\documentclass{pasj01}
\Received{$\langle$reception date$\rangle$}
\Accepted{$\langle$acception date$\rangle$}
\Published{$\langle$publication date$\rangle$}
\usepackage{multirow}
\usepackage{url}
\usepackage[switch,mathlines]{lineno}

\begin{document}

\title{Multiwavelength observation of an active M-dwarf star EV Lac and its stellar flare accompanied by a delayed prominence eruption}
\author{Shun Inoue\altaffilmark{1}\altaffilmark{*}}
\author{Teruaki Enoto\altaffilmark{1,2}}
\author{Kosuke Namekata\altaffilmark{3}}
\author{Yuta Notsu\altaffilmark{4,5}}
\author{Satoshi Honda\altaffilmark{6}}
\author{Hiroyuki Maehara\altaffilmark{7}}
\author{Jiale Zhang\altaffilmark{8}}
\author{Hong-Peng Lu\altaffilmark{8}} 
\author{Hiroyuki Uchida\altaffilmark{1}}
\author{Takeshi Go Tsuru\altaffilmark{1}}
\author{Daisaku Nogami\altaffilmark{9,10}}
\author{Kazunari Shibata\altaffilmark{11,12}}

\altaffiltext{1}{Department of Physics, Kyoto University, Kitashirakawa-Oiwake-cho, Sakyo-ku, Kyoto, 606-8502, Japan}
\altaffiltext{2}{RIKEN Cluster for Pioneering Research, 2-1 Hirosawa, Wako, Saitama, 351-0198, Japan}
\altaffiltext{3}{ALMA Project, NAOJ, NINS, Osawa, Mitaka, Tokyo, 181-8588, Japan}
\altaffiltext{4}{Laboratory for Atmospheric and Space Physics, University of Colorado Boulder, 3665 Discovery Drive, Boulder, CO 80303, USA}
\altaffiltext{5}{National Solar Observatory, 3665 Discovery Drive, Boulder, CO 80303, USA}
\altaffiltext{6}{Nishi-Harima Astronomical Observatory, Center for Astronomy, University of Hyogo, Sayo, Hyogo, 679-5313, Japan}
\altaffiltext{7}{Okayama Branch Office, Subaru Telescope, NAOJ, NINS, Kamogata, Asakuchi, Okayama, 719-0232, Japan}
\altaffiltext{8}{School of Earth and Space Sciences, Peking University, Beijing 100871, China}
\altaffiltext{9}{Department of Astronomy, Kyoto University, Kitashirakawa-Oiwake-cho, Sakyo-ku, Kyoto, 606-8502, Japan}
\altaffiltext{10}{Astronomical Observatory, Kyoto University, Sakyo-ku, Kyoto, 606-8502, Japan}
\altaffiltext{11}{Kwasan Observatory, Kyoto University, Yamashina, Kyoto, 607-8471, Japan}
\altaffiltext{12}{School of Science and Engineering, Doshisha University, Kyotanabe, Kyoto, 610-0321, Japan}
\email{inoue.shun.57c@kyoto-u.jp}


\maketitle
\begin{abstract}
We conducted 4-night multiwavelength observations of an active M-dwarf star EV Lac on 2022 October 24$-$27 with simultaneous coverage of soft X-rays (NICER; 0.2$-$12 $\mathrm{keV}$, Swift XRT; 0.2$-$10 $\mathrm{keV}$), near-ultraviolet (Swift UVOT/UVW2; 1600$-$3500 $\mathrm{\AA}$ ), optical photometry (TESS; 6000$-$10000 $\mathrm{\AA}$), and optical spectroscopy (Nayuta/MALLS; 6350$-$6800 $\mathrm{\AA}$). 
During the campaign, we detected a flare starting at 12:28 UTC on October 25 with its white-light bolometric energy of $3.4 \times 10^{32}$ erg. 
At about 1 hour after this flare peak, our $\mathrm{H\alpha}$ spectrum showed a blue-shifted excess component at its corresponding velocity of $\sim 100 \: \mathrm{km \: s^{-1}}$. 
This may indicate that the prominence erupted with a 1-hour delay of the flare peak. 
Furthermore, the simultaneous 20-second cadence near-ultraviolet and white-light curves show gradual and rapid brightening behaviors during the rising phase at this flare. 
The ratio of flux in NUV to white light at the gradual brightening was $\sim 0.49$, which may suggest that the temperature of the blackbody is low ($< 9000 \: \mathrm{K}$) or the maximum energy flux of a nonthermal 
electron beam is less than $5\times10^{11} \: \mathrm{erg \: cm^{-2} \: s^{-1}}$.
Our simultaneous observations of NUV and white-light flare raise the issue of a simple estimation of UV flux from optical continuum data
by using a blackbody model.
\end{abstract}
% \pagewiselinenumbers

\begin{table*}[t]
\caption{The observation log of EV Lac during our campaign (2022-10-24 to 2022-10-27)}
\begin{center}
\begin{tabular}{ccccc} 
\hline
\multicolumn{1}{c}{Observation}           & Telescope               & Obs. ID     & Obs. Start time (UTC) & Exposure (ks) \\ \hline \hline
X-ray                  & NICER & 5100420101  & 2022-10-25 13:10      & 3.20              \\
                                         &   (0.2$-$12 keV)                      & 5100420102  & 2022-10-26 12:04      & 5.80               \\
                                         &                         & 5100420103  & 2022-10-27 12:50      & 3.91               \\ \cline{2-5} 
                                         & Swift/XRT & 00031397005 & 2022-10-24 12:14      & 1.64                   \\
                                         &         (0.2$-$10 keV)                & 00031397006 & 2022-10-24 15:31      & 0.86                   \\
                                         &                         & 00031397007 & 2022-10-25 12:07      & 1.65                   \\
                                         &                         & 00031397008 & 2022-10-25 15:27      & 0.82                   \\
                                         &                         & 00031397009 & 2022-10-26 12:07      & 0.39                   \\
                                         &                         & 00031397010 & 2022-10-26 13:32      & 1.32                   \\
                                         &                         & 00031397011 & 2022-10-26 15:10      & 0.82                   \\
                                         &                         & 00031397012 & 2022-10-27 11:55      & 0.48                   \\
                                         &                         & 00031397013 & 2022-10-27 13:31      & 1.44                   \\
                                         &                         & 00031397014 & 2022-10-27 15:06      & 0.96                   \\ \hline
NUV                     & Swift/UVOT & 00031397006 & 2022-10-24 15:31      & 0.85                   \\
                                         &      (1600$-$3500 $\mathrm{\AA}$ / UVW2)      & 00031397007 & 2022-10-25 12:08      & 1.66                   \\
                                         &                         & 00031397008 & 2022-10-25 15:27      & 0.82                   \\
                                         &                         & 00031397009 & 2022-10-26 12:07      & 0.46                   \\
                                         &                         & 00031397011 & 2022-10-26 15:10      & 0.82                   \\ \hline
Optical photometry   & TESS                    &    ---\footnotemark[$*$]         &  ---\footnotemark[$*$]              &       ---\footnotemark[$*$]  \\ 
                                      & (6000$-$10000 $\mathrm{\AA}$) & & & \\ \hline
Optical spectroscopy & Nayuta/MALLS &        ---     &         2022-10-24 11:21              &        $0.3 \times 26 + 0.18 \times 81$  \footnotemark[$\dag$]          \\
\multicolumn{1}{l}{}                                      &     (6350$-$6800 $\mathrm{\AA}$)                         &     ---        &         2022-10-25 12:22               &         $0.18 \times 105$\footnotemark[$\dag$]         \\
\multicolumn{1}{l}{}                                      &                              &      ---       &           2022-10-26 11:00             &              $0.18 \times 120$\footnotemark[$\dag$]  \\
\multicolumn{1}{l}{}     &                              &      ---       &          2022-10-27 11:45             &  $0.18 \times 110$\footnotemark[$\dag$] \\ \hline
\end{tabular}
\end{center}
\begin{tabnote}
    \footnotemark[$*$] TESS (Sector 57 in Cycle 5) always observed EV Lac at 20 sec cadence during our campaign.  \\
    \footnotemark[$\dag$] This means “exposure of each frame” $\times$ “the number of frame”. Since the weather at the Nishi-Harima Astronomical Observatory was unstable on October 24, we set the exposure time to 300 seconds for each frame during cloudy conditions.
\end{tabnote}
\label{tab:obs_log}
\end{table*}

\section{Introduction}
The Sun and cool stars suddenly release magnetic energy stored around star spots in the form of flares. 
A flare emits a wide range of radiation from radio waves to X-rays.
Part of magnetic energy is used for plasma ejections called prominence eruptions \citep{Sinha_2019}.
When the velocity of them is sufficiently large, solar prominence eruptions often lead to coronal mass ejections (CMEs) (e.g., \cite{Shibata_2011}).
The extent to which the flare, prominence, and CME relationships that have been established for the Sun are valid for other stars is not fully understood. Therefore, observational studies of them on stars are actively being conducted.

In the last ten years, optical spectroscopic studies have shown that chromospheric lines during stellar flares sometimes show “blueshifts” or “blue asymmetries”, which indicate plasma motion toward us (\cite{Vida_2016}; \cite{Honda_2018}; \cite{Vida_2019}; \cite{Muheki_2020a}; \cite{Muheki_2020b}; \cite{Maehara_2021}; \cite{Lu_2022}; \cite{Namekata_2022a}; \cite{Inoue_2023}; \cite{Notsu_2023}; \cite{Namekata_2023}). 
These results would suggest that CMEs occur in conjunction with flares not only on the Sun but also on other stars (\cite{Leitzinger_2022b}; \cite{Namekata_2022b}). 
Some stellar flares and CMEs are much larger than those of the Sun (e.g., \cite{Inoue_2023}), and their mechanisms are not well understood.
Since there are still few multiwavelength examples of stellar flares, it is imperative to accumulate simultaneous coverage of flares in X-rays, UV, and radio band together with optical detection of blueshifts of chromospheric lines.

One of the main motivations for studying stellar magnetic activity is to evaluate its impact on exoplanets (\cite{Osten_2015}; \cite{Airapetian_2020}). 
X-rays and UV radiation of stellar flares and high-energy particles produced by CMEs affect chemical compositions and escape rates of the atmosphere of exoplanets (e.g., \cite{Airapetian_2016}; \cite{Segura_2018}; \cite{Konings_2022}).  
\citet{Mulkidjanian_2003} also suggests that UV emission gives the selective advantage to the genesis of DNA and RNA of life by modeling the polymerization.
Thus we need to increase our knowledge of the CME characteristics obtained by optical spectroscopy and high-energy observations (X-ray and UV).

Simultaneous observations, especially in near ultraviolet (NUV; $1600-3500 \: \mathrm{\AA}$) and white light ($6000-10000 \: \mathrm{\AA}$), can also contribute to understanding of the flare spectral model.
Though the spectral energy distribution of optical-to-NUV flares has been assumed to be a single-temperature blackbody component at $\sim 10^{4} \: \mathrm{K}$ ($\simeq 0.86 \: \mathrm{eV}$) (\cite{Mochnacki_1980}; \cite{Hawley_1992}), recent NUV/white-light simultaneous observations have shown that this simple blackbody model can not adequately describe the observed spectra (\cite{Kowalski_2013}; \cite{Kowalski_2016}; \cite{Kowalski_2019}; \cite{Brasseur_2023}; \cite{Jackman_2023}). 
Furthermore, some fast (second-scale) optical photometric observations of stellar flares have been conducted (\cite{Kowalski_2016}; \cite{Aizawa_2022}; \cite{Howard_2022}).
The new 20 second cadence mode of Transiting Exoplanet Survey Satellite (TESS; \cite{Ricker_2015}) mission revealed substructures during the rise phase of the white-light flare in many cases \citep{Howard_2022}.
The effect of the UV flare component on exoplanet has been discussed based on only existing optical data (\cite{Feinstein_2020}; \cite{Howard_2020}), whereas the substructure in white-light flares has yet to be investigated in detail compared with the NUV flare.

In this study, we simultaneously performed X-ray observations (0.2$-$12 keV), NUV observations (1600$-$3500 $\mathrm{\AA}$), optical photometric observations (6000$-$10000 $\mathrm{\AA}$) and optical spectroscopic observations (6350$-$6800 $\mathrm{\AA}$) to an active M-dwarf star EV Lac.
The observation and data reduction (Section \ref{sec:observation_reduction}), analysis and results (Section \ref{sec:analysis_results}), and discussion and conclusions (Section \ref{sec:discussion}) on the details of the flare and the associated $\mathrm{H \alpha}$ blueshift obtained through the simultaneous multiwavelength observations are reported in this paper.

\begin{figure*}[t] 
\begin{center}
\includegraphics[width=17.7cm]{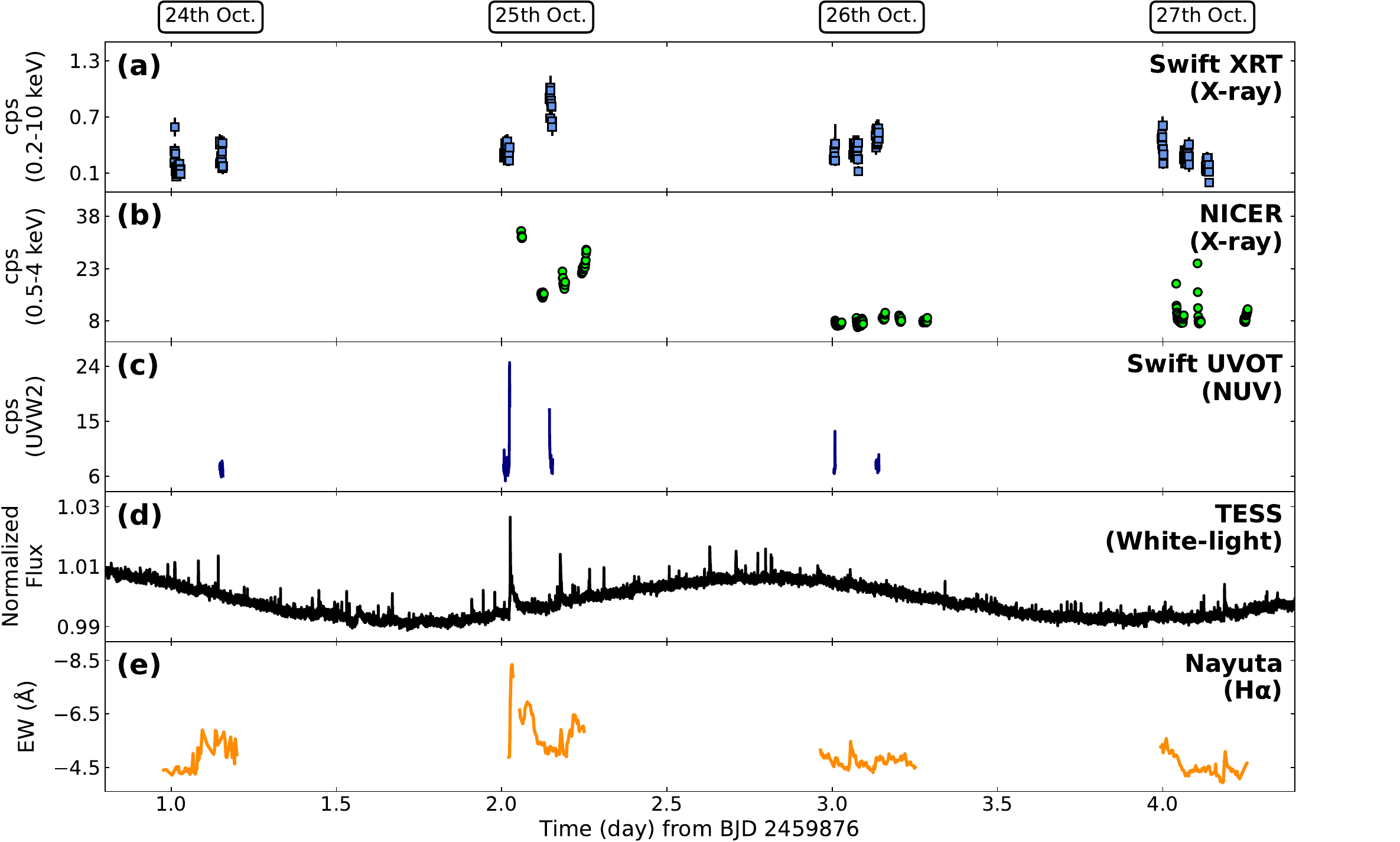} 
\end{center}
\caption{Four-days light curves of EV Lac during our observation campaign on 2022 October 24-27. The time-zero of 2459876 BJD corresponds to 2022 October 23 11:53 UTC. (a) Swift XRT count rates ($\mathrm{counts \: s^{-1}}$) in 0.2$-$10 keV. The time bin and error bars are 64 seconds and one standard deviation statistical error, respectively. (b) NICER count rates ($\mathrm{counts \: s^{-1}}$) in 0.5$-$4 keV. The time bin and error bars are 64 seconds and one standard deviation statistical error, respectively. (c) Swift UVOT count rates ($\mathrm{counts \: s^{-1}}$) in the UVW2 band (1600$-$3500 $\mathrm{\AA}$). The time bin is 20 seconds. (d) TESS white-light light curve shown at 6000$-$10000 $\mathrm{\AA}$. The flux is normalized by the median value. The time bin is 20 seconds. (e) Nayuta $\mathrm{H\alpha}$-line light curve. The equivalent width ($\mathrm{\AA}$) is negative values for the emission line flux in this light curve. The time bin is 180 seconds except for during the part of October 24.
} 
\label{fig:all_days_curve} 
\end{figure*}

\begin{figure*}[t] 
\begin{center}
\includegraphics[width=17.7cm]{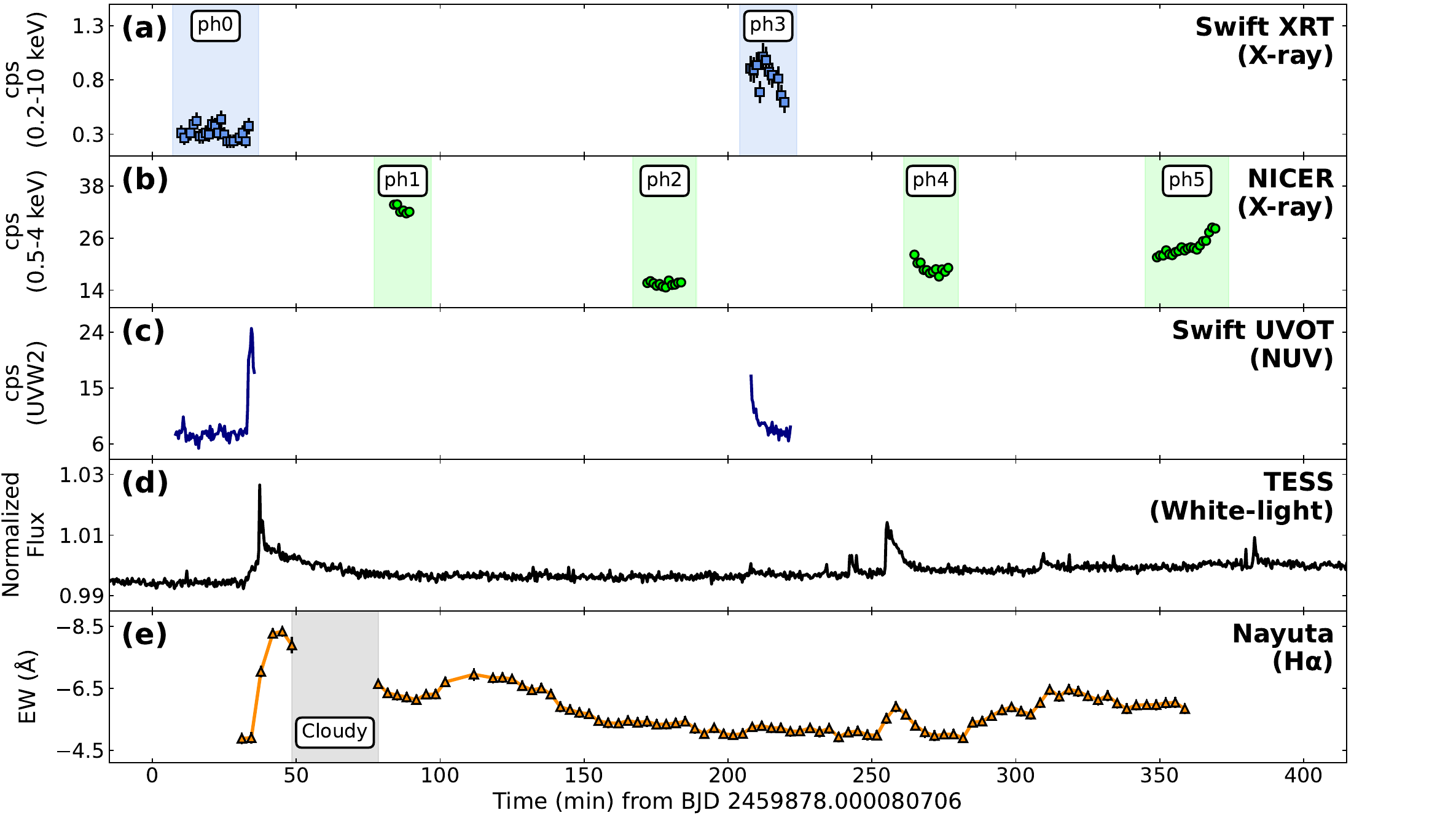} 
\end{center}
\caption{Enlarged light curves of EV Lac on 2022 October 25. The time-zero of 2459878.000080706 BJD corresponds to 2022 October 25 11:52 UTC. Phase numbers in our X-ray spectral analysis are shown as “ph $*$” in panel (a) and (b). The gray zone in panel (e) is cloudy time.
} 
\label{fig:1025_curve} 
\end{figure*}

% \begin{figure}[] 
% \begin{center}
% \includegraphics[width=8.5cm]{./figure/tess_energy_paper_0728.pdf} 
% \end{center}
% \caption{White-light light curves of EV Lac observed with TESS. Time setting is the same as in Figure \ref{fig:1025_curve}. (a) Long term light curve of EV Lac. The flux is normalized by the median value. The vertical green area indicates the flare discussed in this paper. The blue dash-dot line shows the global trend of the stellar rotational modulation fitted with a trigonometric function. (b) Detrended white-light light curve of EV Lac around the flare discussed in this paper. This corresponds to the TESS normalized flux (black line in panel (a)) subtracted by the rotational modulation (blue dash-dot line in panel (a)). The vertical green area is the same as in panel (a).} 
% \label{fig:tess_energy} 
% \end{figure}

\begin{figure*}[t]
\begin{center}
\includegraphics[width=17.3cm]{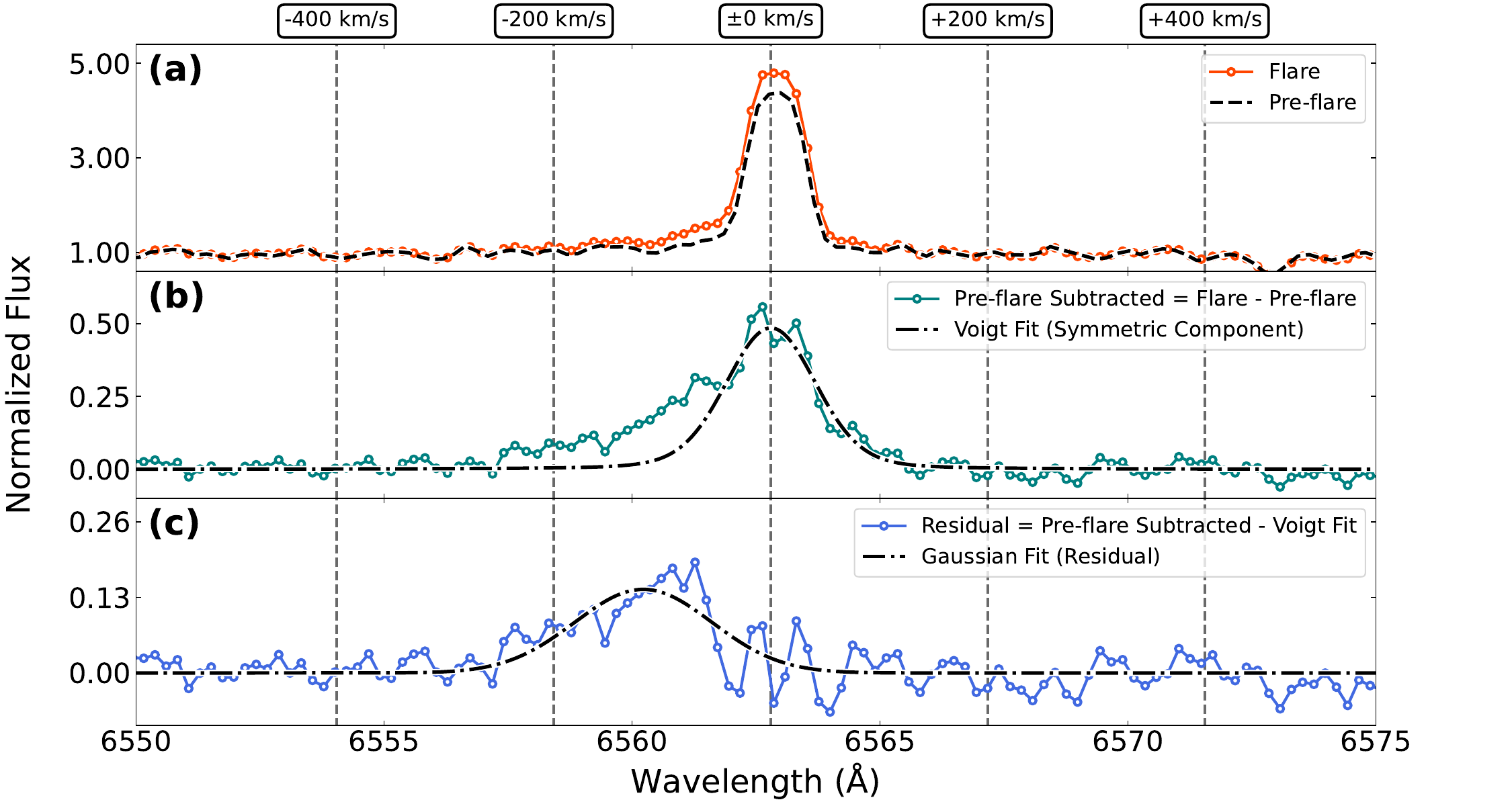} 
\end{center}
\caption{An example of normalized $\mathrm{H\alpha}$ specta extracted from the time period of 127$-$129 min in the light curve of Figure \ref{fig:1025_curve}. Gray vertical dashed lines show Doppler velocity from the line center at $6562.8 \: \mathrm{\AA}$. (a) Comparison of the continuum at the 127$-$129 min range (orange solid line) with that in the pre-flare data at the 30$-$35 min (black dashed line). (b) Pre-flare subtracted spectrum of panel (a). The black dashdot line represents the Voigt function fitting only used for the red side of the line center (6562.8$-$6600 $\mathrm{\AA}$). (c) Residual between the pre-flare subtracted spectrum (panel b) and the best-fit Voigt function result in panel (b). The black dashdot line represents the Gaussian fitting.}
\label{fig:Ha_spectrum} 
\end{figure*}

\begin{figure}[b!]
\begin{center}
\includegraphics[width=7.6cm]{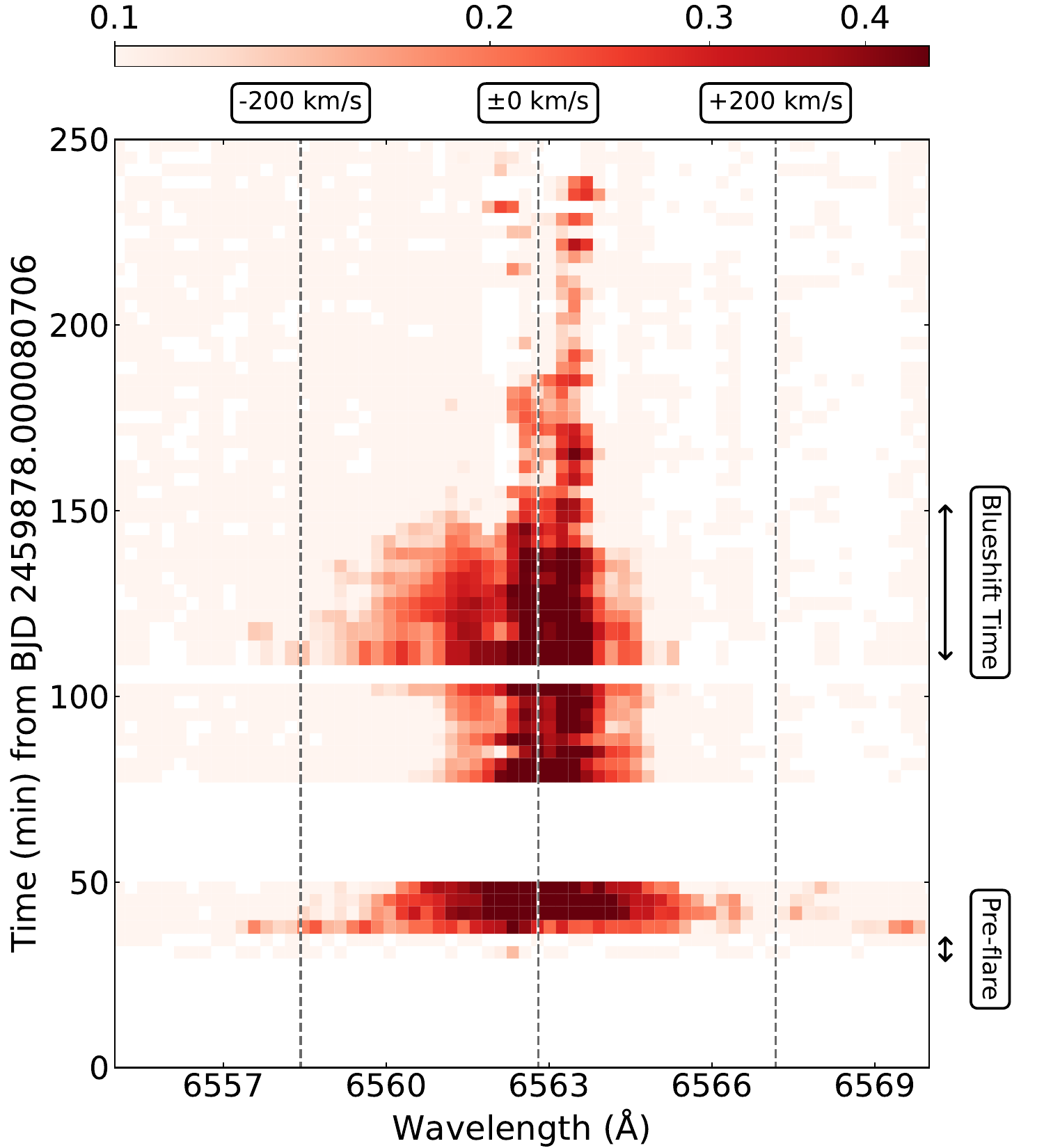} 
\end{center}
\caption{Time variation of pre-flare subtracted spectrum. The horizontal and vertical  axes  represent the wavelength and time, respectively. The time origin is the same as Figure \ref{fig:1025_curve}. The red color scale indicates flux normalized by the continuum. Gray dashed lines show Doppler velocity from the line center ($6562.8 \: \mathrm{\AA}$).}
\label{fig:Ha_2d_spectrum} 
\end{figure}

\begin{figure*}[] 
\begin{center}
\includegraphics[width=16cm]{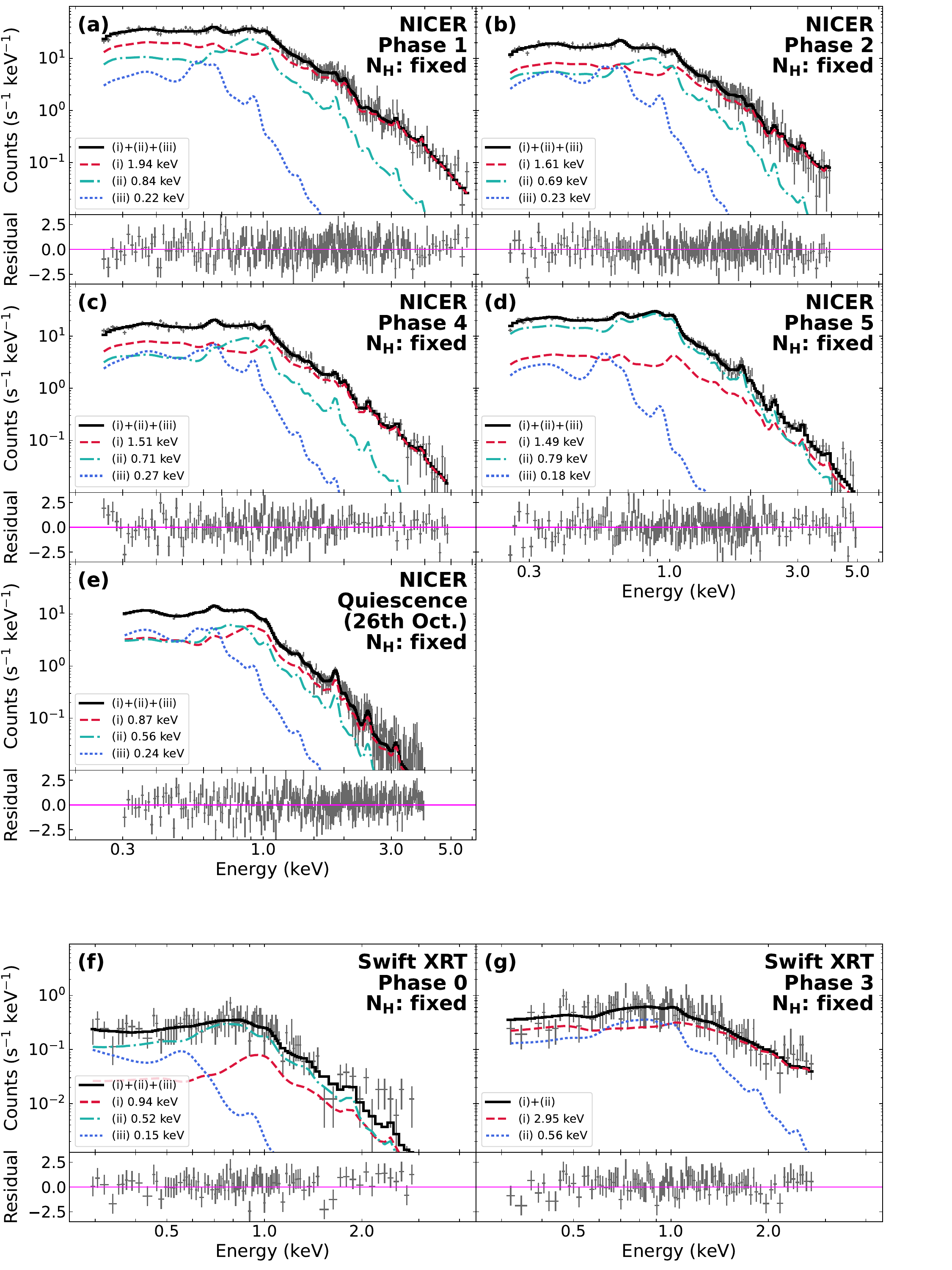} 
\end{center}
\caption{Background-subtracted and response-uncorrected X-ray spectra at each phase shown in Figure \ref{fig:1025_curve}. The best-fit curves for three-temperature \texttt{vapec} models are shown by solid black lines. Red dashed, green dashdot, and blue dotted lines represent the high, medium, and low temperature component, respectively. Panels (a-d) and (g) are spectra during the flare, whereas panels (e) and (f) are spectra during the quiescence.}
\label{fig:all_X_spectrum_app} 
\end{figure*}

\section{Observation and Data Reduction} \label{sec:observation_reduction}
\subsection{Target Star}
EV Lac (GJ 873) is an M4.5 Ve star. Many stellar flares have been detected from EV Lac with X-ray (e.g., \cite{Favata_2000}) and optical (e.g., \cite{Honda_2018}) observations. 
Some studies investigated the flare frequency of EV Lac (\cite{Muheki_2020b}; \cite{Paudel_2021}; \cite{Ikuta_2023}).
Table 1 in \citet{Paudel_2021} summarizes the basic physical parameters of EV Lac. 
The distance to EV Lac is 5.05 pc \citep{Gaia_2018}, and we use this value throughout in our paper.
\citet{Honda_2018}, \citet{Muheki_2020b}, and \citet{Notsu_2023} discovered a $\sim 100 \: \mathrm{km \: s^{-1}}$ blueshift of $\mathrm{H \alpha}$ line on EV Lac.
Blueshifts of X-ray lines on EV Lac, which may be attributed to prominence eruptions or chromospheric evaporation, are also reported in \citet{Chen_2022}.
Since these observations were only in one band, the total energy distribution of the flare and prominence eruption was unknown.

\subsection{Multiwavelength Observations}
We conducted a 4-day multiwavelength (X-ray, NUV, optical photometry, and optical spectroscopy) observation campaign of EV Lac on 2022 October 24$-$27. Observation times of all telescopes are summarized in Table \ref{tab:obs_log}.
\subsubsection{NICER: X-ray}
We used NASA's Neutron Star Interior Composition ExploreR (NICER; \cite{Gendreau_2016}) to conduct soft X-ray ($0.2-12$ keV) observations of EV Lac via our request to a ToO program. 
NICER performed monitoring on EV Lac for three days from October 25 to 27 (See Table \ref{tab:obs_log}).
On each day, NICER made 3$-$5 observations, of which exposure time was $\sim 1$ ks each.

We retrieved observation data from HEASARC Archive.
We employed the standard data analysis procedure of NICER.
At first, we used \texttt{nicerl2} in HEASoft ver. 6.30.1 to filter and calibrate raw data using the calibration database (\texttt{CALDB}) version \texttt{xti20221001}. Filtered data were barycenter corrected using  \texttt{barycorr} at the target position of (\texttt{RA, DEC}) = (341.707214, 44.333993). Then, we extracted light curves from the filtered and barycenteric-corrected event file with \texttt{xselect}. We also generated source and background spectra with \texttt{nibackgen3C50} \citep{Remillard_2022}. We produced the response files, i.e., \texttt{RMF} and \texttt{ARF} files, using \texttt{nicerrmf} and \texttt{nicerarf} commands, respectively. \texttt{XSPEC ver. 12.12.1} \citep{Arnaud_1996} was used for our spectral analysis.
\subsubsection{Swift XRT: X-ray}
NASA's Neil Gehrels Swift Observatory (Swift) also observed EV Lac in soft X-ray ($0.2-10$ keV) on October 24$-$27 via our ToO request.
Swift performed 2$-$3 observations per day with the X-ray telescope (XRT; \cite{Burrows_2005}) in Photon Counting (PC) mode.

Data reduction was conducted in the same manner as that in \citet{Paudel_2021} with \texttt{barycorr}, \texttt{xrtpipeline}, \texttt{xselect}.
The source and background region were set to a 30-pixels (71 arcsec) radius circle and an annular extreaction region with inner and outer radi of 40 (94 arcsec) and 70 (165 arcsec) pixels, respectively. Both region's center was set to the position of the source. We used \texttt{swxpc0to12s6\_20130101v014.rmf} available in the \texttt{CALDB} file for the response file.
We produced \texttt{ARF} files using \texttt{xrtmkarf}.
Spectral analysis was also conducted by using \texttt{XSPEC} ver. 12.12.1 \citep{Arnaud_1996}.
\subsubsection{Swift UVOT: NUV}
Swift UVOT also observed EV Lac with UVW2 filter during almost the same time period as the XRT observation. In some observations (Obs-IDs 00031397005, 00031397010, 00031397012, 00031397013, and 00031397014), EV Lac was located outside the field of view of UVOT. UVW2 filter passes NUV light. The central wavelength and FWHM of UVW2 filter are $1928 \: \mathrm{\AA}$ and $657 \: \mathrm{\AA}$, respectively \citep{Poole_2008}.

Data reduction was conducted in the same manner as that in \citet{Paudel_2021} with \texttt{coordinator}, \texttt{uvotscreen}, \texttt{barycorr}, and \texttt{uvotevtlc}. 
Only for the Obs-ID 00031397007, we set \texttt{uvotscreen} filtering as \texttt{evexpr=(QUALITY\%256).eq.0} after confirming with the Swift helpdesk because the vast majority of events indicate \texttt{QUALITY=256}.
When we created the light curve with \texttt{uvotevtlc}, time bin (\texttt{timedel}) was set to 20 seconds, the same as that of the TESS light curve.
\subsubsection{TESS: Optical Photometry}
TESS conducted optical photometric observations of EV Lac in Sector 57 during Cycle 5 for one month (2022 September 30$-$October 29) with the 20 second cadence. The band of TESS filter is 6000$-$10000 $\mathrm{\AA}$. We downloaded the observation data from the MAST data archive at the Space Telescope Science Institute (STScI).
We read and analyzed the Pre-search Data Conditioning Simple Aperture Photometry (\texttt{PDCSAP flux}) in the data by using \texttt{python} ver. 3.9.12 and \texttt{astropy} ver. 5.0.4.
\subsubsection{Nayuta: Optical Spectroscopy}
We conducted optical spectroscopic observations of EV Lac with Nayuta. Nayuta is 2m telescope at the Nishi-Harima Astronomical Observatory in Japan. We used a spectrometer of Medium And Low-dispersion Long-slit Spectrograph (MALLS), whose wavelength resolution ($R = \lambda / \Delta \lambda$) is $\sim 10000$ at 6500 $\mathrm{\AA}$. MALLS covers 6350$-$6800 $\mathrm{\AA}$ including the $\mathrm{H \alpha}$ line. During most of the time, the exposure time of each frame was set to 180 seconds, but on the first day (October 24), it was occasionally increased to 300 s due to unstable weather conditions.

Data processing was conducted in the same manner as that in \citet{Honda_2018} using \texttt{IRAF}\footnote{IRAF is distributed by the National Optical Astronomy Observatories, which are operated by the Association of Universities for Research in Astronomy, Inc., under cooperate agreement with the National Science Foundation.} package \citep{Today_1986}.
We made a minor adjustment to the wavelength calibration by using photospheric line profiles after the \texttt{IRAF} processing.

\begin{table*}[]
\caption{Best-fit parameters of all spectra with three temperature collisionally-ionized models\footnotemark[$*$]}
\label{tab:fitting_3vapec}
\begin{center}
\begin{tabular}{ccccccccc}
\hline
\multicolumn{2}{c}{\multirow{2}{*}{}} & \multicolumn{5}{c}{NICER} & \multicolumn{2}{c}{Swift XRT} \\
\multicolumn{2}{c}{} & Phase 1 & Phase 2 & Phase 4 & Phase 5 & Oct 26 & Phase 0 & Phase 3 \\ \hline \hline
\multicolumn{2}{c}{Exposure (ks)} & $0.42$ & $0.71$ & $0.71$ & $1.19$ & $5.80$ & $1.65$ & $0.82$  \\ \hline
\multicolumn{2}{c}{\texttt{tbabs}} & & & & & & & \\
\multicolumn{2}{c}{$N_{\mathrm{H}}$ ($10^{18}$ $\mathrm{cm^{-2}}$)} & 4.0 & 4.0 & 4.0 & 4.0 & 4.0 & 4.0 & 4.0 \\ \hline
\multicolumn{2}{c}{\texttt{vapec} (High Temp.)} & & & & & & & \\
\multirow{2}{*}{Temperature} & $kT$ (keV) & $1.94_{\pm0.08}$ & $1.61_{\pm0.06}$ & $1.51_{\pm0.06}$ & $1.49_{\pm0.09}$ & $0.87_{\pm0.04}$ & $0.94_{\pm0.35}$ & $2.95_{\pm1.42}$ \\
& $T$ (MK) & $22.5_{\pm0.9}$ & $18.8_{\pm0.7}$ & $17.6_{\pm0.7}$ & $17.3_{\pm1.0}$ & $10.1_{\pm0.5}$ & $10.9_{\pm4.1}$ &  $34.2_{\pm16.5}$\\
\multicolumn{2}{c}{norm ($10^{-2}$)} & $2.60_{\pm0.11}$ & $0.94_{\pm0.05}$ & $0.88_{\pm0.04}$ & $0.50_{\pm0.07}$ & $0.49_{\pm0.10}$ & $0.11_{\pm0.15}$ & $1.20_{\pm0.16}$ \\ \hline
\multicolumn{2}{c}{\texttt{vapec} (Medium Temp.)} & & & & & & & \\
\multirow{2}{*}{Temperature} & $kT$ (keV) & $0.84_{\pm0.02}$ & $0.69_{\pm0.03}$ & $0.71_{\pm0.04}$ & $0.79_{\pm0.01}$ & $0.56_{\pm0.05}$ & $0.52_{\pm0.14}$ & --- \\
& $T$ (MK) & $9.74_{\pm0.23}$ & $8.00_{\pm0.35}$ & $8.24_{\pm0.46}$ & $9.16_{\pm0.12}$ & $6.50_{\pm0.58}$ & $6.03_{\pm1.62}$ & --- \\
\multicolumn{2}{c}{norm ($10^{-2}$)} & $1.15_{\pm0.16}$ & $0.55_{\pm0.09}$ & $0.42_{\pm0.08}$ & $1.58_{\pm0.10}$ & $0.43_{\pm0.07}$ & $0.43_{\pm0.16}$ & --- \\ \hline
\multicolumn{2}{c}{\texttt{vapec} (Low Temp.)} & & & & & & & \\
\multirow{2}{*}{Temperature} & $kT$ (keV) & $0.22_{\pm0.03}$ & $0.23_{\pm0.01}$ & $0.27_{\pm0.02}$ & $0.18_{\pm0.03}$ & $0.24_{\pm0.01}$ & $0.15_{\pm0.10}$ & $0.56_{\pm0.13}$\\
& $T$ (MK) & $2.55_{\pm0.35}$ & $2.67_{\pm0.12}$ & $3.13_{\pm0.23}$ & $2.09_{\pm0.35}$ & $2.78_{\pm0.12}$ & $1.74_{\pm1.16}$ & $6.20_{\pm1.44}$   \\
\multicolumn{2}{c}{norm ($10^{-2}$)} & $0.24_{\pm0.09}$ & $0.21_{\pm0.04}$ & $0.21_{\pm0.04}$ & $0.10_{\pm0.02}$ & $0.26_{\pm0.03}$ & $0.15_{\pm0.10}$ & $0.52_{\pm0.12}$ \\ \hline
\multirow{13}{*}{$Z$ ($Z_{\odot}$)} & He & $1.00$ & $1.00$ & $1.00$ & $1.00$ & $1.00$ & $1.00$ & $1.00$ \\
& C & $0.59_{\pm0.09}$ & $0.52_{\pm0.06}$ & $0.53_{\pm0.05}$ & $0.63_{\pm0.08}$ & $0.44_{\pm0.02}$ & $0.44$ & $0.52$ \\
& N & $0.59_{\pm0.09}$ & $0.52_{\pm0.06}$ & $0.53_{\pm0.05}$ & $0.63_{\pm0.08}$ & $0.44_{\pm0.02}$ & $0.44$ & $0.52$ \\
& O & $0.59_{\pm0.09}$ & $0.52_{\pm0.06}$ & $0.53_{\pm0.05}$ & $0.63_{\pm0.08}$ & $0.44_{\pm0.02}$ & $0.44$ & $0.52$ \\
& Ne & $1.00$ & $1.00$ & $1.28_{\pm0.27}$ & $1.00$ & $0.45_{\pm0.07}$ & $0.45$ & $1.00$ \\
& Mg & $0.23_{\pm0.13}$ & $0.41_{\pm0.12}$ & $0.34_{\pm0.12}$ & $0.29_{\pm0.05}$ & $0.24_{\pm0.04}$ & $0.24$ & $0.41$ \\
& Al & $1.00$ & $1.00$ & $1.00$ & $1.00$ & $1.00$ & $1.00$ & $1.00$ \\
& Si & $0.48_{\pm0.10}$ & $0.30_{\pm0.09}$ & $0.50_{\pm0.10}$ & $0.34_{\pm0.04}$ & $0.40_{\pm0.04}$ & $0.40$ & $0.30$ \\
& S & $0.15_{\pm0.14}$ & $0.41_{\pm0.16}$ & $0.59_{\pm0.16}$ & $0.42_{\pm0.10}$ & $0.46_{\pm0.09}$ & $0.46$ & $0.41$ \\
& Ar & $1.00$ & $1.00$ & $1.00$ & $1.00$ & $1.00$ & $1.00$ & $1.00$ \\
& Ca & $1.00$ & $1.00$ & $1.00$ & $1.00$ & $1.00$ & $1.00$ & $1.00$ \\
& Fe & $0.21_{\pm0.04}$ & $0.13_{\pm0.03}$ & $0.18_{\pm0.03}$ & $0.13_{\pm0.01}$ & $0.14_{\pm0.01}$ & $0.14$ & $0.13$ \\
& Ni & $1.00$ & $1.00$ & $1.00$ & $1.00$ & $1.00$ & $1.00$ & $1.00$ \\ \hline
\multicolumn{2}{c}{$\chi^{2}$ ($d.o.f$)} & 252 (247) & 192 (216) & 194 (156) & 232 (200) & 290 (259) & 71 (87) & 87 (107)\\ \hline
\end{tabular}
\end{center}
\begin{tabnote}
    \footnotemark[$*$] The error ranges correspond to $90\%$ confidence level. Values without errors mean that they are fixed. \\
    % For Phase 3 spectrum, elemental abundances were fixed to the best-fit value of 26th Oct.
\end{tabnote}
\end{table*}

\section{Analysis and Results}\label{sec:analysis_results}
\subsection{Light Curves}\label{sec:light_curve}
Multiwavelength light curves of EV Lac obtained by our campaign are shown in Figure \ref{fig:all_days_curve}.  
A large flare, which is the focus of this paper, occurred on October 25.
The white-light light curve shows that many small flares occur everyday on EV Lac. 

Figure \ref{fig:1025_curve} shows enlarged light curves on October 25. 
We numbered each observation of Swift/XRT and NICER as “Phase (ph) $*$”.
As shown in Figure \ref{fig:1025_curve}, the largest flare, which is the focus of this paper, started at 12:28 UTC on 2023 October 25, corresponding to $\sim 40$ min from the time origin of Figure \ref{fig:1025_curve}. We have succeeded in observing the rising phase of the flare in NUV, white light, and $\mathrm{H\alpha}$.
At the end of Phase 0, UV and white light were already increasing, but X-ray rising was not observed.
Other small flares also occurred, e.g., the X-ray and UV flare at $\sim 210$ min (Phase 3), and the white-light and $\mathrm{H\alpha}$ recorded flare at $\sim 250$ min slightly before Phase 4.
Though there is no large flare in the white-light light curve before and after Phase 5, emission of X-ray and $\mathrm{H\alpha}$ is gradually increasing after $\sim 300 \: \mathrm{min}$. 
Compared to white light, other wavelength emission has greater flare contrast. In addition, there are flares where emission is observed only in X-ray and UV (e.g., the X-ray and UV flare at $\sim 210$ min).
Hereafter, we focus on the largest flare during our observation campaign.

\begin{figure}[t] 
\begin{center}
\includegraphics[width=8.2cm]{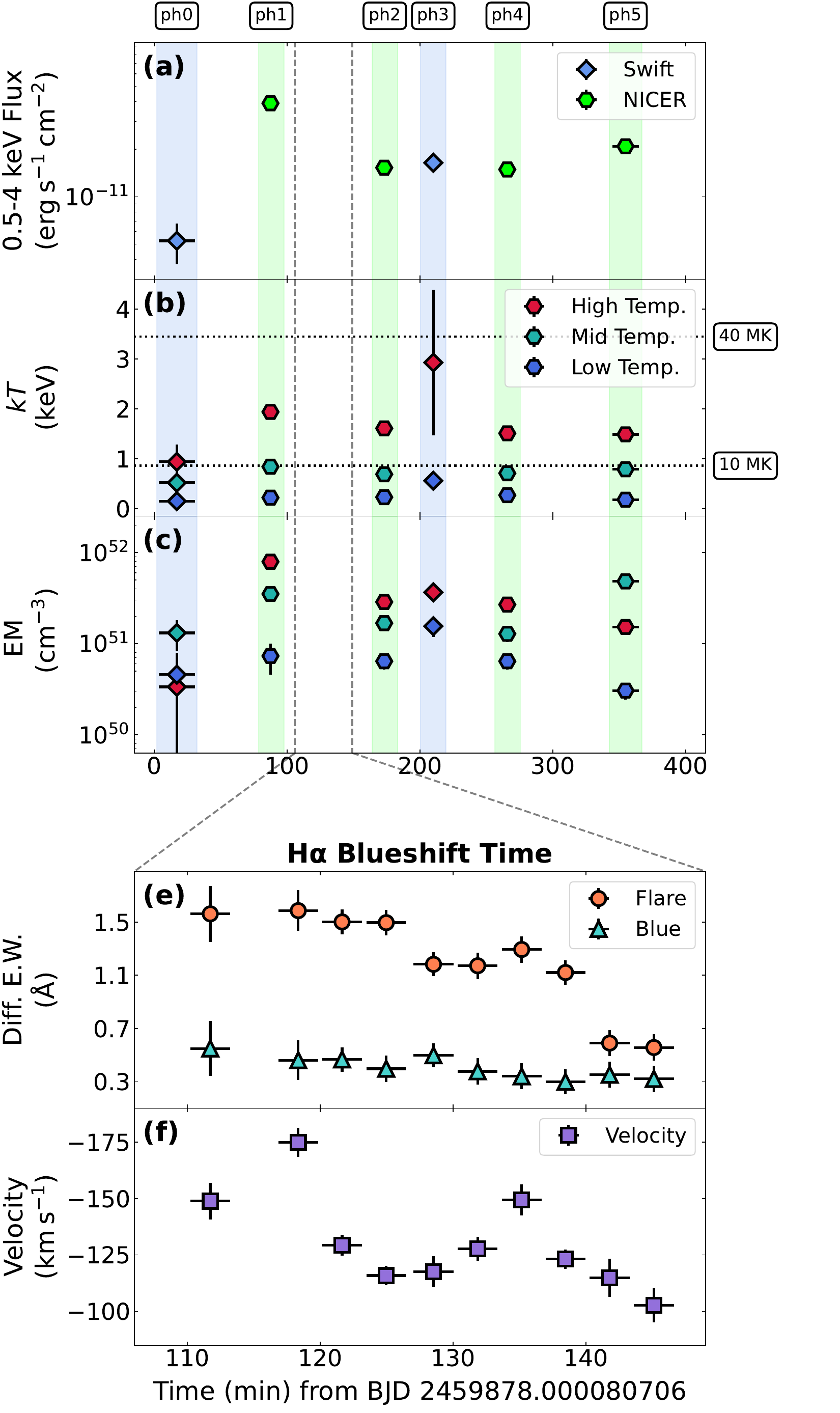} 
\end{center}
\caption{Time evolution of physical parameters obtained by the X-ray and $\mathrm{H\alpha}$ spectral analysis. (a) Time variation of 0.5--4 keV X-ray flux. Blue diamonds and lime hexagons indicate the 0.5--4 keV flux calculated from Swift and NICER X-ray spectra, respectively. Phase numbers are also shown in the same manner as in Figure \ref{fig:1025_curve}. The area sandwiched between gray horizontal lines corresponds to the time period when $\mathrm{H\alpha}$ line was blue-shifted. (b) Temperatures of plasma. Red, green, and blue markers indicate high, medium, and low temperature components, respectively. Diamonds and hexagons corresponds Swift and NICER data, respectively. (c) Emission measure with colors and markers the same as in panel (b). (e) Light curves of the equivalent width of $\mathrm{H\alpha}$ during it was blueshifted. The orange circles and turquoise triangles indicate the equivalent widths of the flare symmetric and blueshifted excess components, respectively. (f) Corresponding velocity of the blueshifted excess components of $\mathrm{H\alpha}$.}
\label{fig:fitting_results} 
\end{figure}

\subsection{Spectral Analysis} \label{sec:ha_X_fitting}
\subsubsection{H$\alpha$} \label{sec:ha_spectral_analysis}
We searched the largest flare for an asymmetric component of $\mathrm{H\alpha}$ in the same manner as \citet{Inoue_2023}.
Figure \ref{fig:Ha_spectrum} shows the $\mathrm{H\alpha}$ spectrum extracted from a frame at 127-129 min (Figure \ref{fig:1025_curve}). At first, we created the “pre-flare” spectrum shown as black dashed line in Figure \ref{fig:Ha_spectrum}a by combining two frames just before the flare start. Then, subtracting the pre-flare spectrum from the flare spectrum (127$-$129 min), we made  pre-flare-subtracted spectrum (Figure \ref{fig:Ha_spectrum}b). Since a blue-shifted excess component was confirmed (Figure \ref{fig:Ha_spectrum}b) at $\sim 6560 \: \mathrm{\AA}$ ($-100 \: \mathrm{km \: s^{-1}}$), we separated it from a symmetric component by fitting only the red side of $\mathrm{H\alpha}$ with the Voigt function.
The line center of the Voigt function  was fixed to the $\mathrm{H\alpha}$ line center (6562.8 $\mathrm{\AA}$).
Figure \ref{fig:Ha_spectrum}c shows the residual between the pre-flare subtracted spectrum and the Voigt function fitted only on the red side. 
Finally, we fit the residual with Gaussian. 
We conducted this Voigt fitting of the symmetric components for all frames on October 25.
We also performed the Gaussian fitting of the blue-shifted excess components for 10 frames at 112$-$145 min, when they are clearly present.
The line center and standard deviation ($\sigma$) of the Gaussian fitted on the blue-shifted excess components were 6558$-$6560 $\mathrm{\AA}$ ($-200 \sim -100$ $\mathrm{km \: s^{-1}}$) and $\sim 1 \: \mathrm{\AA}$ ($ \sim 45$ $\mathrm{km \: s^{-1}}$), respectively.

Figure \ref{fig:Ha_2d_spectrum} shows the time variation of the pre-flare subtracted spectrum. As shown in 110$-$150 min in Figure \ref{fig:Ha_2d_spectrum}, there was a continuous blueshifted excess component in the $\mathrm{H\alpha}$ emission line. The blue-shifted excess component appeared one hour after the flare peak. It coincides with the secondary peak at 112 min of the $\mathrm{H\alpha}$ light curve (see Figure \ref{fig:1025_curve}e).
\subsubsection{X-ray} \label{sec:fitting_X-ray}
We show in Figure \ref{fig:all_X_spectrum_app} the NICER and Swift X-ray spectra of each phase on October 25. We also extracted the NICER spectrum on October 26 as a quiescent data (Figure \ref{fig:all_X_spectrum_app}e), since there is no X-ray flare observed during this period.
The Phase 0 spectrum obtained by Swift XRT is also a quiescent spectrum because the X-ray flare started after the end of Phase 0.

In order to investigate the time variation of temperature, abundance, and emission measure during the flare, we performed X-ray spectral fitting for all the spectra utilizing with three temperature collisionally-ionized equilibrium components (\texttt{vapec}) with interstellar absorption (\texttt{tbabs}). 
Note that although we also tried a single- or two-temperature \texttt{vapec} model, they give a statistically unacceptable fit.
Only for the Phase 3 spectrum, we used two-temperature \texttt{vapec} model because the best-fit parameter of temperature with three-temperature \texttt{vapec} model was physically unacceptable.
We linked abundance among the three components and further tied the abundance of C, N and O, which have the similar first ionization potential.
We fixed the hydrogen column density at $N_{\mathrm{H}} = 4.0 \times 10^{18} \: \mathrm{cm}^{-2}$ \citep{Paudel_2021}, since the small distance, 5.05 pc \citep{Gaia_2018}, to the source prevents us from determining the small interstellar absorption from the observed data.
Due to a lack of statistics, it was difficult to determine abundance of Phase 0 and 3 spectra obtained by Swift XRT. Therefore, abundances of Phase 0 and 3 were fixed to the best-fit value of October 26 NICER spectrum and Phase 2, respectively. 
Figure \ref{fig:all_X_spectrum_app} and Table \ref{tab:fitting_3vapec} summarizes the spectral fitting and its best-fit parameters, respectively.

As an alternative spectral modeling, we also conducted fitting with the fixed quiescent component and two temperature collisionally-ionized models (\texttt{vapec}) with interstellar absorption (\texttt{tbabs}) for flare spectra (Phase 1-5). In the fitting, we could not fit Phase 2-4 spectra with fixed $N_{\mathrm{H}}$. 
Since the best-fit parameters of temperature were physically unacceptable in this modeling, we did not adopt this modeling.
See the Appendix. for more information about the fitting with the fixed quiescent component.

\subsubsection{Time evolution of fitting parameters} \label{sec:fitting_results}
Figure \ref{fig:fitting_results} shows the time evolution of physical parameters obtained by our X-ray and $\mathrm{H\alpha}$ spectral analysis. 
From Phase 1 to Phase 2, the plasma temperature and emission measure of the large flare is cooling and decreasing as indicated in Figure \ref{fig:fitting_results}a$-$c. As the light curve in Figure \ref{fig:1025_curve} also showed, flux, temperature, and emission measure remained higher than those in pre-flare (Phase 0) even after Phase 2 because some other small flares occurred. 
This suggests that these small flares injected energy into the plasma.

As shown in Figure \ref{fig:fitting_results}e, we investigated time variation of the equivalent widths of $\mathrm{H\alpha}$ line, and decomposed them into the flare symmetric component and the blueshifted excess component. We also calculated the Doppler velocity of the blueshifted excess component (Figure \ref{fig:fitting_results}f). 
The equivalent widths of the flare and blueshifted excess components were calculated by integrating Voigt function (cf. Figure \ref{fig:Ha_spectrum}b) and Gaussian (cf. Figure \ref{fig:Ha_spectrum}c), respectively.
The Doppler velocity of the blueshifted excess component corresponds to the wavelength of the center of Gaussian (cf. Figure \ref{fig:Ha_spectrum}c).
The equivalent width of the blueshifted excess component was $\sim 1/3$ of the flare symmetric component at $\sim 110 \: \mathrm{min}$. 
Then, both components were decaying and the difference became progressively smaller.
Time variation of the Doppler velocity of the blueshifted excess component appeasrs to have two peaks at $\sim 120$ and $\sim 135 \:\mathrm{min}$. These velocities of $100-200 \: \mathrm{km \: s^{-1}}$ are comparable to the blueshift on mid M dwarf stars (\cite{Honda_2018}; \cite{Vida_2019}; \cite{Notsu_2023}).

 \begin{figure}[] 
\begin{center}
\includegraphics[width=7.8cm]{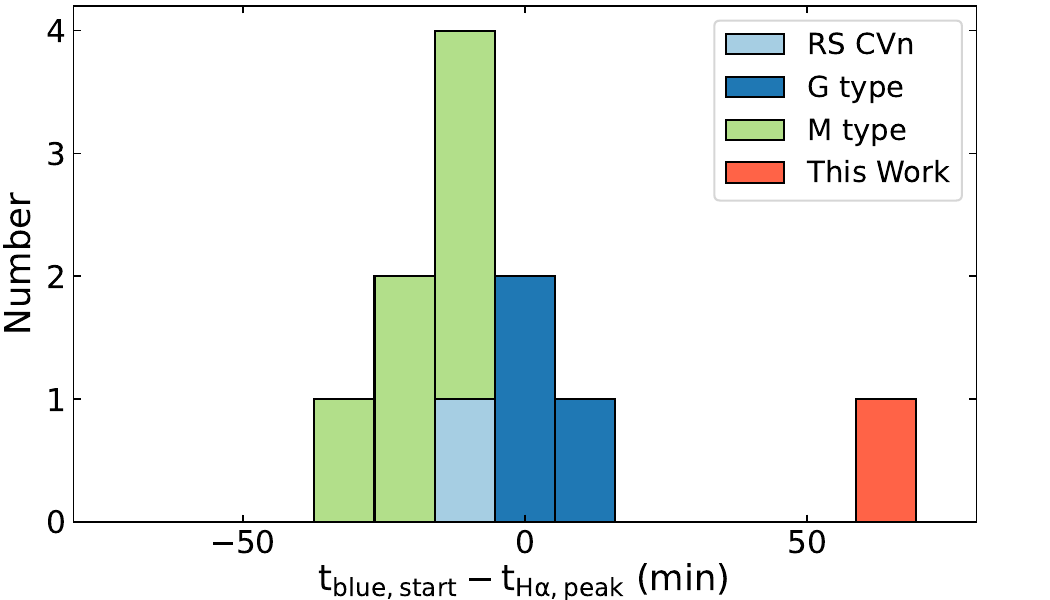} 
\end{center}
\caption{Distribution of the difference between the appearance time of $\mathrm{H\alpha}$ blue-shifted components ($t_{\mathrm{blue, start}}$) and flare peak ($t_{\mathrm{H\alpha, peak}}$). Light blue, blue and green correspond to $\mathrm{H\alpha}$ blue-shift events on RS CVn (\cite{Inoue_2023}), G (\cite{Namekata_2022a}; \cite{Namekata_2023}), and M (\cite{Vida_2016}; \cite{Honda_2018}; \cite{Maehara_2021}; \cite{Notsu_2023}) type stars, respectively. Orange corresponds to this work.} 
\label{fig:blue_hist} 
\end{figure}

\begin{figure*}[t] 
\begin{center}
\includegraphics[width=17cm]{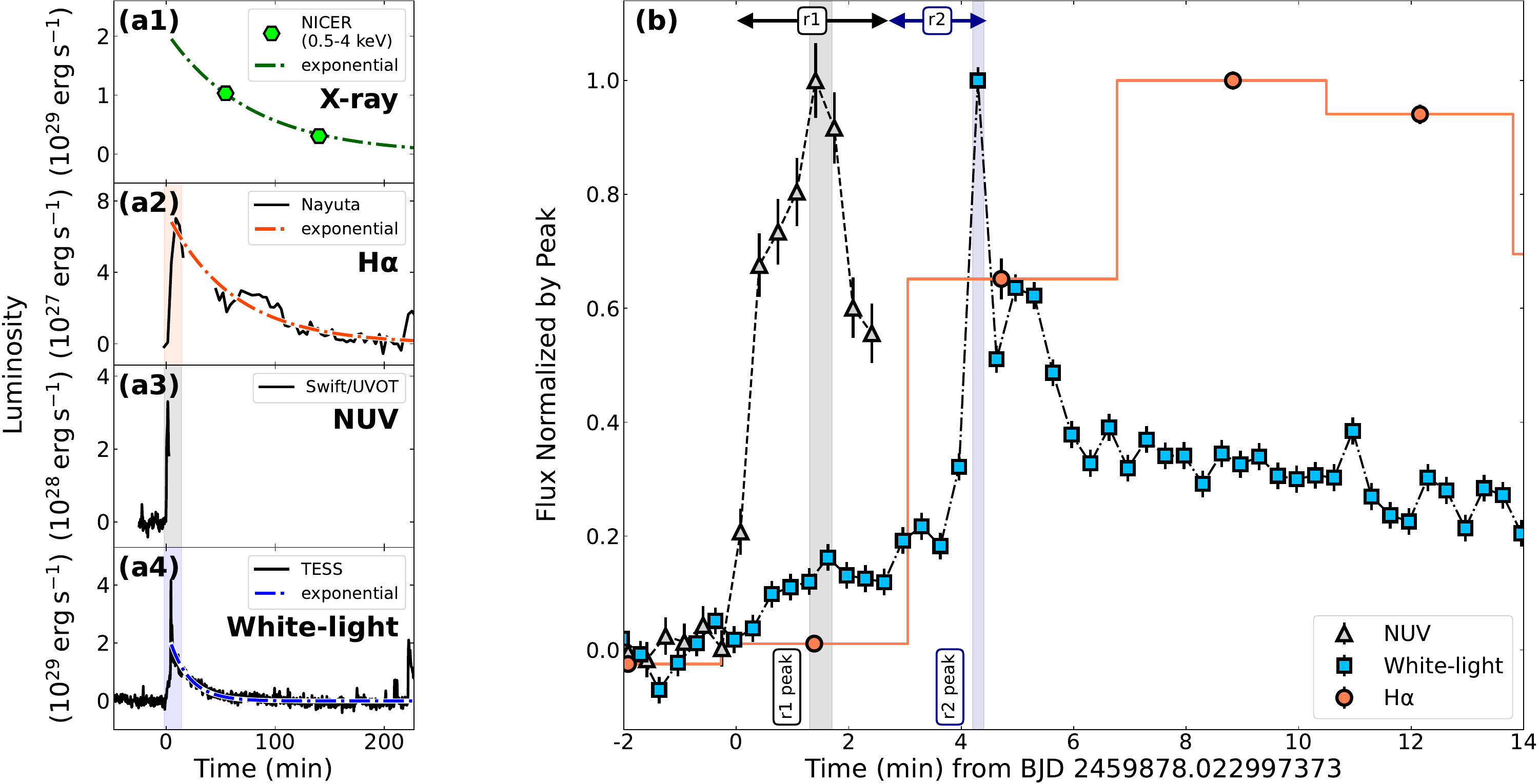} 
 \end{center}
\caption{(a1-a4) Quiescent-subtracted light curves. The time zero point is set to the beginning of the flare. The dashdot lines show the exponential function fitted for the decay phase of the light curve. (a1) Flare luminosity of 0.5$-$4 keV X-ray. (a2) Flare luminosity of $\mathrm{H\alpha}$. Note that the luminosity of the blue-shifted excess component is removed in this curve. (a3) Flare luminosity of UVW2 band (1600$-$3500 $\mathrm{\AA}$). (a4) Flare luminosity of whight light (6000$-$10000 $\mathrm{\AA}$). (b) Enlarged light curves around the rising phase of the flare, which corresponds to the shaded time intervals in panel (a2$-$a4). All curves are normalized by their peak values. Aqua triangle, blue square, and orange circle represent NUV, white-light, and $\mathrm{H\alpha}$, respectively. Blue and green shaded regions indicates gradual and rapid phase of white-light, respectively.}
\label{fig:rising_curve} 
\end{figure*}

\section{Discussion}\label{sec:discussion}
\subsection{Blue-shifts and prominence eruptions}
The $\mathrm{H\alpha}$ blue shifted excess components were identified during the flare (Section \ref{sec:ha_spectral_analysis}, Figure \ref{fig:Ha_spectrum} \& \ref{fig:Ha_2d_spectrum}).
The stellar rotational velocity ($v \sin i \sim 3.5$ $\mathrm{km \: s^{-1}}$; \cite{Reiners_2018}) can not explain the observed blueshifts ($\sim 100$ $\mathrm{km \: s^{-1}}$).

There are two candidates for such moving plasma. The one is a chromospheric temperature (cool) upflow associated with chromospheric evaporation (\cite{Tei_2018}) and the other is a prominence eruption (\cite{Otsu_2022}).
Since we colud not observe signs of the mass ejection in X-ray and UV, such as the coronal dimming (\cite{Veronig_2021}; \cite{Loyd_2022}) and the increase of the hydrogen density (\cite{Moschou_2017}; \cite{Moschou_2019}), we can not assure that the observed blue-shifts are attributed to the prominence eruption.
However, the velocity of cool upflow ($\sim 10$ $\mathrm{km \: s^{-1}}$; \cite{Tei_2018}) is typically smaller than the observed velocity ($\sim 100 \: \mathrm{km \: s^{-1}}$) in the case of solar flares. Furthermore, there was no significant increase in white light when the blue-shifts appeared. 
According to numerical simulations, the injection of non-thermal electrons into deep chrmosphere can produce unheated cool upward flows above the  chromospheric evaporation. Such non-thermal electrons are also expected to produce chromospheric condensation, producing significant white-light emissions (e.g., \cite{Li_2023}). 
The lack of white-light emission therefore may indicate that the above process is not working.

\citet{Honda_2018} also discussed the possibility of the absorption by post-flare loops making the blue asymmetry during the decay phase.
However, the $\mathrm{H\alpha}$ spectra in this study did not show a sharp red-shifted absorption as observed by \citet{Honda_2018}.

For these reasons, it is highly probable that the  prominence eruption occurred and made the observed blue-shifted excess components.

 \subsubsection{Timing of the prominence eruption}
 One interesting point in the present event is the timing of $\mathrm{H\alpha}$ blue-shift. In many cases, blue-shifted excess components of $\mathrm{H\alpha}$ are pronounced at the flare peak (Figure \ref{fig:blue_hist})\footnote{For the flare described in \citet{Vida_2016}, we took the difference between $t_{1}$ and the flare peak (cf. Figure 14 in \cite{Vida_2016}) because blue-shifts at $t_{2}$ and $t_{3}$ may be bulue-shifts without flares (\cite{Muheki_2020b}). For the flare Y6 in \citet{Notsu_2023}, we considered time [1] in Figure 14 in \citet{Notsu_2023} to be the flare peak because there were multiple peaks during the flare, which suggests that some flares occurred simultaneously.}. In other words, most prominence eruptions are initiated at the flare peak. However, the prominence discovered in this study appeared $\sim 1$ hour after the peak of the flare.
There are three possible cases for interpreting the delay of the $\mathrm{H\alpha}$ blue-shift. 

First, another flare occurred $\sim 1$ hour after the peak of the first flare. There is no obvious signs of another flare in the white-light (Figure \ref{fig:1025_curve}d), while there are tiny enhancements in $\mathrm{H\alpha}$ line center emission (Figure \ref{fig:Ha_2d_spectrum}). It is possible that a non-white light flare accompanied by a mass ejection occurred.
It is also possible that the prominence erupted with no flare connections. Some studies have reported the prominence eruption without obvious flare connections on the Sun \citep{Zirin_1969, Mason_2021}. 

Second, the prominence erupted during the decay phase of the flare. 
\citet{Kurokawa_1987} reported an X13 class solar flare in which a filament erupted about 40 minutes later than the major flare peak.
The change of magnetic field configuration due to the flare reconnection should make a twisted filament start to be ejected and accelerated by magnetic force in a twisted tube \citep{Shibata_1986}.
Though such delayed prominence eruptions have not been observed on stars (Figure \ref{fig:blue_hist}), this interpretation is consistent with our observations.

% \color{red}
% Third, the prominence erupted with no flare connections. Some studies have reported the prominence eruption without obvious flare connections on the Sun, which can possibly represent a failed CME or coronal rain \citep{Zirin_1969, Mason_2021}.
% \citet{Mason_2021} discussed the possibility that such eruptions could be caused by interchange reconnection.
% \color{black}

Third, the prominence erupted on the disk at the flare peak and outside the limb $\sim 1$ hour after the flare peak. Generally, prominence eruptions on disk and outside a limb are observed as $\mathrm{H\alpha}$ absorption and emission, respectively in the case of the Sun and solar-type stars (\cite{Parenti_2014}; \cite{Otsu_2022}; \cite{Namekata_2022a}; \cite{Namekata_2023}). It is possible that the prominence initially erupted on disk and erupted outside a limb in the course of the one-hour trip. If we assume the prominence visibility on M-dwarf is the same as those on the Sun and solar-type stars, this possibility is unlikely because $\mathrm{H\alpha}$ spectra between the flare peak and the start of blue-shifted emission components show no signs of blue-shifted absorption components. However, \citet{Leitzinger_2022} estimated that prominence on disk can be $\mathrm{H\alpha}$ emission for dM stars using 1D NLTE modeling and cloud model formulation, so the source function of $\mathrm{H\alpha}$ line of M-dwarf prominences could be relatively comparable to or higher than background continuum radiation. Therefore, in some specific prominence parameters, there could be a possibility that the erupted prominence is initially invisible inside the disk and appeared as an emission after going outside the stellar limb. 
Since we do not know the source function of M-dwarf prominence, we need to perform simultaneous observations of some Balmer lines (c.f. \cite{Vida_2016}; \cite{Notsu_2023}) in the future to verify this interpretation.
Furthermore, given that the prominence moved $ \sim 3.6 \times 10^{5} \: \mathrm{km}$ ($\sim 1.5 R_{\rm star}$) and continued to expanding over $\sim$ one hour, it is unclear whether it retains enough emission measure to be observed as $\mathrm{H \alpha}$ emission at the time.
Theoretical simulations are needed to investigate the time variation of emission measure of the prominence.

\subsection{Physical parameters} \label{sec:physical_par}
From the results of spectral analyse in Section \ref{sec:ha_X_fitting}, we calculated basic physical parameters of the flare and prominence.

\subsubsection{Prominence Mass}
We estimated the mass of the prominence using a method used by \citet{Maehara_2021}, \citet{Inoue_2023}, and \citet{Notsu_2023}. 
As shown in Figure \ref{fig:fitting_results}e, the maximum equivalent width of the blueshifted excess component is $\sim 0.5 \: \mathrm{\AA}$.
The equation (5) in \citet{Notsu_2023} presented the formula to convert the equivalent width ($EW_{\mathrm{H\alpha}}$) of $\mathrm{H\alpha}$ to its luminosity,
\begin{equation}
\label{eq:L_EW}
    L_{\mathrm{H \alpha}} (t) = 4 \pi d^2 \times F^{\mathrm{cont}}_{\mathrm{H\alpha}} \times EW_{\mathrm{H\alpha}} (t),
\end{equation}
where $F^{\mathrm{cont}}_{\mathrm{H\alpha}}$ ($5.7 \times 10^{-13} \: \mathrm{erg \: cm^{-2} \: s^{-1} \: \AA^{-1}}$; \cite{Notsu_2023}) is the quiescent flux density at the continuum level around $\mathrm{H\alpha}$ of EV Lac, and $d$ ($5.05 \: \mathrm{pc}$; \cite{Gaia_2018}) is the distance between the Earth and the target star. 
Using Equation (\ref{eq:L_EW}), we calculated the luminosity of the blueshifted excess component $L_{\mathrm{blue}}$:
\begin{equation}
\label{eq:L_blue}
    L_{\mathrm{blue}} \sim 8 \times 10^{26} \; \mathrm{erg \: s^{-1}}.
\end{equation}
We adopted the non-local thermodynamic equilibrium (NLTE) model of the solar prominence \citep{Heinzel_1994} and the range of the optical thickness $\tau_{\mathrm{p}}$ of $\mathrm{H\alpha}$ line center is assumed to be $0.1<\tau_{\mathrm{p}}<100$ as done in \citet{Inoue_2023}.
\begin{enumerate}
    \item $\tau_{\mathrm{p}} \sim 0.1$: NLTE model \citep{Heinzel_1994} indicates that the $\mathrm{H\alpha}$ flux of the  prominence per unit time, unit area, and unit solid angle $F_{\mathrm{H \alpha}}$ is
    \begin{equation}
    \label{eq:F_value} 
    F_{\mathrm{H\alpha}} \sim 10^{4} \; \mathrm{erg \: s^{-1} \: cm^{-2} \: sr^{-1}}. 
    \end{equation}
    As shown in Equation (8) in \citet{Inoue_2023}, $L_{\mathrm{blue}}$ is expressed as
    \begin{equation}
    \label{eq:L_blue2} 
    L_{\mathrm{blue}} = \int \int F_{\mathrm{H\alpha}} \: dA d\Omega = 2 \pi A_{\mathrm{p}} F_{\mathrm{H\alpha}},
    \end{equation}
    where $A_{\mathrm{p}}$ is the area of the region emitting $\mathrm{H\alpha}$. Using Equations (\ref{eq:L_blue})$-$(\ref{eq:L_blue2}), we obtained
    \begin{equation}
    \label{eq:A_value} 
    A_{\mathrm{p}} \sim 1 \times 10^{22} \; \mathrm{cm^{2}} \sim 6 \times A_{\mathrm{star}}, 
    \end{equation}
    where $A_{\mathrm{star}} = \pi R_{\mathrm{star}}^{2}$ ($\sim 2\times10^{21} \: \mathrm{cm}^{2}$) is the area of the hemisphere of the star and $R_{\mathrm{star}}$ ($\sim 3 \times 10^{10} \: \mathrm{cm}$; \cite{Paudel_2021}) is the radius of the star.
    \citet{Heinzel_1994} also indicates that in the case of Equation (\ref{eq:F_value}), the emission measure $n_{\mathrm{e}}^{2}D$ of the prominence is 
    \begin{equation} 
    n_{\mathrm{e}}^{2} D \sim 10^{28} \; \mathrm{cm^{-5}} \label{eq:EM_value}, 
    \end{equation}
    where $D$ and $n_{\mathrm{e}}$ are the geometrical thickness and the electron density of the prominence, respectively.
    \citet{Hirayama_1986} shows the typical electron density of a solar prominence is
    \begin{equation}
    10^{10} \; \mathrm{cm^{-3}} < n_{\mathrm{e}} < 10^{11.5} \; \mathrm{cm^{-3}} \label{eq:ne_value}.
    \end{equation}
    \citet{Notsu_2023} calculated the ratio between the hydrogen density $n_{\mathrm{H}}$ and the electron density $n_{\mathrm{e}}$ of a prominence,
    \begin{equation}
    n_{\mathrm{e}} / n_{\mathrm{H}} = 0.17-0.47 \label{eq:ne_nh_ratio}
    \end{equation}
    from Table 1 of \citet{Labrosse_2010}. The mass of the prominence is expressed as 
    \begin{equation}
    M \sim m_{\mathrm{H}} n_{\mathrm{H}} A_{\mathrm{p}} D, \label{eq:mass_eq}
    \end{equation}
    where $m_{\mathrm{H}}$ is the mass of hydrogen atom. From Equations (\ref{eq:A_value})-(\ref{eq:mass_eq}), 
    \begin{equation} 
    \label{eq:M_value_01}
    1 \times 10^{15} \; \mathrm{g} < M < 1 \times 10^{17} \; \mathrm{g}
    \end{equation} 
    is obtained.
    \item $\tau_{\mathrm{p}} \sim 100$: Calculated as in case $\tau_{\mathrm{p}} \sim 0.1$, 
    \begin{equation}
    F_{\mathrm{H\alpha}} \sim 10^{6} \; \mathrm{erg \: s^{-1} \: cm^{-2} \: sr^{-1}}, 
    \end{equation}
    \begin{equation}
    \label{eq:A_value_100}
    A_{\mathrm{p}} \sim 1 \times 10^{20} \; \mathrm{cm^{2}} \sim 6 \times 10^{-2} \times A_{\mathrm{star}},
    \end{equation}
    \begin{equation} 
    \label{eq:ne2D31}
    n_{\mathrm{e}}^{2} D \sim 10^{31} \; \mathrm{cm^{-5}}, 
    \end{equation}
    \begin{equation} 
    \label{eq:M_value_100}
    1 \times 10^{16} \; \mathrm{g} < M < 1 \times 10^{18} \; \mathrm{g}.
    \end{equation}     
\end{enumerate}
% From Equations (\ref{eq:A_value}) and (\ref{eq:A_value_100}), we obtained the range of $A_{\mathrm{p}}$,
% \begin{equation}
%     6 \times 10^{-2} \times A_{\mathrm{star}} < A_{\mathrm{p}} < 6 \times A_{\mathrm{star}}.
% \end{equation}
We obtained the range of $M$ from Equations (\ref{eq:M_value_01}) and (\ref{eq:M_value_100}),
\begin{equation} 
\label{eq:M_value_comb_range}
1 \times 10^{15} \; \mathrm{g} < M < 1 \times 10^{18} \; \mathrm{g}.
\end{equation} 
This mass and the white-light bolometric flare energy $E_{\mathrm{WLF, bol}}$ of $3.4\times10^{32} \: \mathrm{erg}$ (see Section \ref{sec:energy distribution}) are comparable to previous blueshifts on M-dwarf stars (\cite{Moschou_2019}; \cite{Maehara_2021}; \cite{Notsu_2023}) and correspond to the value expected from the flare energy-mass scaling law $M \propto E_{\mathrm{WLF, bol}}^{2/3}$ (\cite{Takahashi_2016}; \cite{Namekata_2022a} \cite{Inoue_2023}; \cite{Namekata_2023}).

\begin{table}[b]
\caption{Flare loop size and magnetic field strength estimated from X-ray spectrum of phase1}
\begin{center}
\begin{tabular}{c|ccc}
\hline
 $n_{0}$  & $B_{\mathrm{SY}}$ & $L_{\mathrm{SY}}$ & $L_{\mathrm{N}}$  \\ \hline \hline
$10^{11} \: \mathrm{cm^{-3}}$ & $\sim 80 \: \mathrm{G}$  & $\sim 0.7 R_{\mathrm{star}}$   &  $\sim 0.2 R_{\mathrm{star}}$  \\
$10^{12} \: \mathrm{cm^{-3}}$ & $\sim 160 \: \mathrm{G}$  & $\sim 0.3 R_{\mathrm{star}}$  &  $\sim 0.1 R_{\mathrm{star}}$  \\
$10^{13} \: \mathrm{cm^{-3}}$ & $\sim 300 \: \mathrm{G}$  & $\sim 0.1 R_{\mathrm{star}}$  &  $\sim 0.09 R_{\mathrm{star}}$\\ 
\hline
\end{tabular}
\end{center}
\label{tab:L_and_B}
\end{table}

\subsubsection{Flare Loop Size}
\citet{Shibata_2002} showed magnetic reconnection model equations for calculating the length of a flare loop $L_{\mathrm{SY}}$ and the flare magnetic field  strength $B_{\mathrm{SY}}$,
\begin{eqnarray} 
L_{\mathrm{SY}} = &10^{9}& \left( \frac{EM_{\mathrm{peak}}}{10^{48} \: \mathrm{cm}^{-3}} \right)^{3/5} \nonumber \\ &\times& \left( \frac{\mathnormal{n_{0}}}{10^{9} \: \mathrm{cm}^{-3}} \right)^{-2/5} \left( \frac{\mathnormal{T_{\mathrm{peak}}}}{10^{7} \: \mathrm{K}} \right)^{-8/5}
\: \mathrm{cm}, \label{eq:Shibata_Yokoyama_L}
\end{eqnarray}
\begin{eqnarray}
B_{\mathrm{SY}} = &50& \left( \frac{EM_{\mathrm{peak}}}{10^{48} \: \mathrm{cm}^{-3}} \right)^{-1/5} \nonumber \\ &\times& \left( \frac{\mathnormal{n_{0}}}{10^{9} \: \mathrm{cm}^{-3}} \right)^{3/10} \left( \frac{\mathnormal{T_{\mathrm{peak}}}}{10^{7} \: \mathrm{K}} \right)^{17/10}
\: \mathrm{G},  \label{eq:Shibata_Yokoyama_B}
\end{eqnarray}
where $EM_{\mathrm{peak}}$ is the peak volume emission measure, $T_{\mathrm{peak}}$ is the peak temperature, and $n_{0}$ is the preflare coronal density. \citet{Osten_2006} placed a constraint on the coronal electron density of EV Lac between $10^{10} \: \mathrm{cm}^{-3}$ and $10^{14} \: \mathrm{cm}^{-3}$ using X-ray and UV density-sensitive line ratios. For coronal temperature during our quiescent phase (Table \ref{tab:fitting_3vapec}), coronal density is assumed to be $10^{11} - 10^{13} \: \mathrm{cm}^{-3}$ (see Figure 9 in \cite{Osten_2006}). We substituted temperature and emission measure obtained from the X-ray spectrum of Phase 1, which is closest to the flare peak, for Equations (\ref{eq:Shibata_Yokoyama_L}) and (\ref{eq:Shibata_Yokoyama_B}). 
As a result, the flare magnetic field  strength and loop size are $80\:\mathrm{G} < B_{\mathrm{SY}} < 300\:\mathrm{G}$ and $0.1 R_{\mathrm{star}} < L_{\mathrm{SY}} < 0.7 R_{\mathrm{star}}$, respectively.

We also calculated the flare loop size $L_{\mathrm{N}}$ by using the equation derived by \citet{Namekata_2017b} and \citet{Namekata_2023}:
\begin{eqnarray}
    L_{\mathrm{N}} &=& 1.64 \times 10^{9} \left(\frac{\tau_{\mathrm{decay}}^{\mathrm{WLF}}}{100 \: \mathrm{s}} \right)^{2/5} \nonumber \\ 
    &\times& \left( \frac{E_{\mathrm{WLF, bol}}}{10^{30} \: \mathrm{erg}} \right)^{1/5} \left( \frac{n_{0}}{n_{\odot}}\right)^{-1/5} \: \mathrm{cm},
\end{eqnarray}
where $\tau_{\mathrm{decay}}^{\mathrm{WLF}}$ is the $e$-folding time of the white-light flare. The calculated value is $0.09 R_{\mathrm{star}} < L_{\mathrm{N}} < 0.2 R_{\mathrm{star}}$
Table \ref{tab:L_and_B} compiles the results of our calculation. 
$L_{\mathrm{SY}}$ and $L_{\mathrm{N}}$ are the almost same order for each coronal density.

\begin{table*}[]
\caption{Peak luminosity and duration of the flare at each wavelength}
\begin{center}
\begin{tabular}{ccccc}
\hline
  & X-ray & NUV & White Light & $\mathrm{H\alpha}$  \\
  & $0.5-4 \: \mathrm{keV}$ & $1600-3500 \: \mathrm{\AA}$ & $6000-10000 \: \mathrm{\AA}$ & $6562.8 \: \mathrm{\AA}$ \\ \hline \hline
Peak luminosity ($10^{29} \: \mathrm{erg \: s^{-1}}$) & $1.9_{-0.1}^{+0.2}$ & $0.33_{\pm 0.02}$ \footnotemark[$*$] &  $0.67_{\pm 0.1}$ \footnotemark[$*$] / $4.2_{\pm 0.1}$ \footnotemark[$\dag$] & $0.07_{\pm 0.001}$ \\
Rising time (min) & $-$ & $1.7_{\pm 0.1}$ \footnotemark[$*$] & $7.0_{\pm 0.1}$ & $7.4_{\pm 1.9}$ \\
e-Folding time (min) & $77_{\pm 9.6}$ & $-$ & $19_{\pm1.2}$ & $61_{\pm 7.6}$ \\
  \hline 
\end{tabular}
\end{center}
\begin{tabnote}
    \footnotemark[$*$]These values are at r1 peak. \\ 
    \footnotemark[$\dag$]This value is at r2 peak.
\end{tabnote}
\label{tab:flare_parameters}
\end{table*}

\begin{table*}[]
\caption{Energy of radiation at each wavelength and the mass ejection}
\begin{center}
\begin{tabular}{cccccc}
\hline
\multicolumn{5}{c}{Radiation}  & Mass Ejection \\ \hline
X-ray & NUV & \multicolumn{2}{c}{White Light} & $\mathrm{H\alpha}$ & Kinetic Energy \\
 $0.5-4 \: \mathrm{keV}$ & $1600-3500 \: \mathrm{\AA}$ & Bolometric & 6000$-$10000 $\mathrm{\AA}$ & $6562.8 \: \mathrm{\AA}$ & \\
  ($\mathrm{erg}$) & ($\mathrm{erg}$) & ($\mathrm{erg}$) & ($\mathrm{erg}$) & ($\mathrm{erg}$) & ($\mathrm{erg}$) \\ \hline \hline
   $9.2_{-1.6}^{+2.3} \times 10^{32}$   &   $(0.2-4.0) \times 10^{31}$  &    $3.4_{\pm1.1} \times 10^{32}$ &   $2.4_{\pm1.2} \times 10^{32}$   &  $2.7_{\pm0.4} \times 10^{31}$  &    $(0.02-15) \times 10^{31}$            \\ \hline            
\end{tabular}
\end{center}
\label{tab:distribution}
\end{table*}

\subsection{Multiwavelength Rising Phase Data}\label{sec:rising}
Figure \ref{fig:rising_curve}b shows the enlarged light curve of the rising phase of the flare. 
Similar to solar flares, white light and NUV due to non-thermal emission increases faster than $\mathrm{H\alpha}$, which is called Neupert effect (e.g., \cite{Neupert_1968}; \cite{Namekata_2020}; \cite{Tristan_2023}).

The rising phase of the white-light flare consists of two phases (Figure \ref{fig:rising_curve}b): a gradual rise (r1) and a rapid rise (r2). \citet{Howard_2022} showed many samples of flares which exhibit a similar substructure using 20 second cadence mode data of TESS. Our new finding in this study is that there is already a sharp increase in NUV during the white-light gradual phase (r1). The last three bins of the NUV light curve appear to have already begun to decay.
Since the NUV observations stopped $\sim 35 \: \mathrm{min}$ in Figure \ref{fig:rising_curve}, it is not clear whether the NUV flux rose sharply again or continued to decay during r2.

The ratio of NUV flux to white-light flux is crucial to the model of the broadband spectrum of stellar flares (e.g., \cite{Jackman_2023}; \cite{Brasseur_2023}). However, there are few simultaneous observations of stellar flares in NUV and white-light band. 
While we missed NUV flux during r2, we calculated the ratio of NUV flux to white-light one at the r1 peak.

\begin{figure}[] 
\begin{center}
\includegraphics[width=8.5cm]{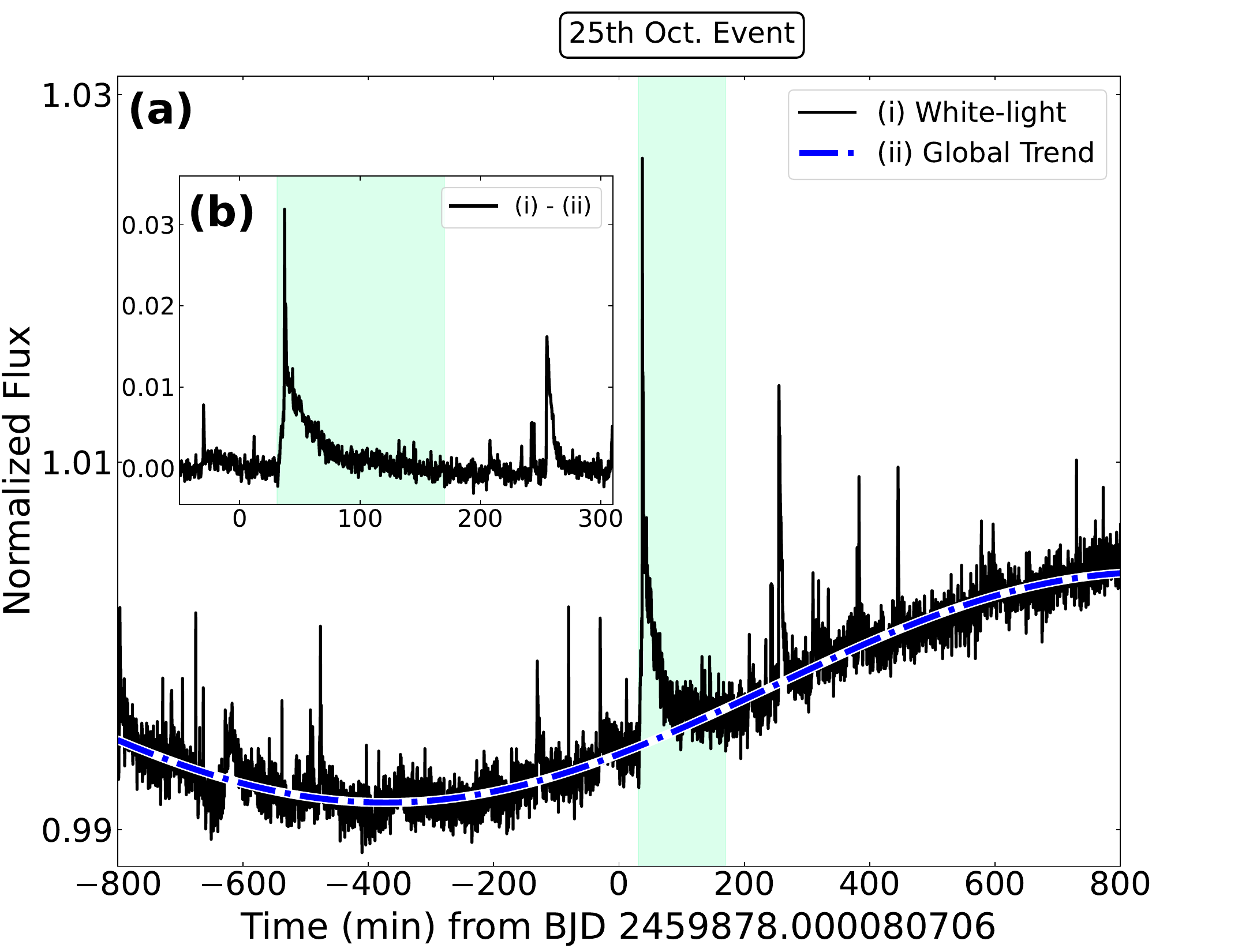} 
\end{center}
\caption{White-light light curves of EV Lac observed with TESS. Time setting is the same as in Figure \ref{fig:1025_curve}. (a) Long-term curve of EV Lac. The flux is normalized by the median value. The vertical green area indicates the flare discussed in this paper. The blue dash-dot line shows the global trend of the stellar rotational modulation fitted with a trigonometric function. (b) Detrended white-light light curve of EV Lac around the flare discussed in this paper after subtracting of the rotational modulation.} 
\label{fig:tess_energy} 
\end{figure}

We evaluated the luminosity at the r1 peak in the TESS band ($L^{\mathrm{TESS}}_{\mathrm{peak}}$) in the same manner as \citet{Notsu_2023}:
\begin{equation}
\label{eq:L_TESS}
    L^{\mathrm{TESS}}_{\mathrm{peak}} = L^{\mathrm{TESS}}_{\mathrm{quiescence}} \times \Delta f^{\mathrm{TESS}}_{\mathrm{peak}} = 6.7 \times 10^{28} \: \mathrm{erg \: s^{-1}},
\end{equation}
where $L^{\mathrm{TESS}}_{\mathrm{quiescence}}$ ($1.3 \times 10^{31} \: \mathrm{erg \: s^{-1}}$; \cite{Notsu_2023}) is the quiescent luminosity of EV Lac in TESS band and $\Delta f^{\mathrm{TESS}}_{\mathrm{peak}} = (f_{\mathrm{TESS, p}} - f_{\mathrm{TESS, q}}) / f_{\mathrm{TESS, q}}$ is the relative flux at the flare peak (cf. Figure \ref{fig:tess_energy}b). 
$f_{\mathrm{TESS, p}}$ and $f_{\mathrm{TESS, q}}$ are TESS flux at the r1 peak and of quiescence, respectively. See Section \ref{sec:energy distribution} and Figure \ref{fig:tess_energy} for more information about the TESS flux. 
We also calculated the luminosity at the flare peak in UVW2 band ($L^{\mathrm{UVW2}}_{\mathrm{peak}}$) from the value of \texttt{FLUX\_AA} in the light curve file created by \texttt{uvotevtlc}:
\begin{eqnarray}
\label{eq:L_UVW2}
    L^{\mathrm{UVW2}}_{\mathrm{peak}} &=& (F^{\mathrm{UVW2}}_{\mathrm{peak}} - F^{\mathrm{UVW2}}_{\mathrm{quiescence}}) \times W^{\mathrm{UVW2}}_{\mathrm{eff}} \times 4 \pi d^{2}  \nonumber \\ 
    &=&  3.3 \times 10^{28} \: \mathrm{erg \: s^{-1}},
\end{eqnarray}
where $F^{\mathrm{UVW2}}_{\mathrm{peak}}$ and $F^{\mathrm{UVW2}}_{\mathrm{quiescence}}$ are flux density (\texttt{FLUX\_AA}) at the flare peak and pre-flare, respectively. $W^{\mathrm{UVW2}}_{\mathrm{eff}} $ ($667.73 \: \mathrm{\AA}$; SVO Filter Profile Service\footnote{\url{http://svo2.cab.inta-csic.es/theory/fps/index.php?id=Swift/UVOT.UVW2&&mode=browse&gname=Swift&gname2=UVOT#filter}}) is the equivalent width of the effective area of UVW2 filter, and $d$ ($5.05 \: \mathrm{pc}$; \cite{Gaia_2018}) is the distance between the Earth and EV Lac.
From Equations (\ref{eq:L_TESS}) and (\ref{eq:L_UVW2}), 
\begin{equation}
     L^{\mathrm{UVW2}}_{\mathrm{peak}} / L^{\mathrm{TESS}}_{\mathrm{peak}} \sim 0.49
\end{equation}
is obtained. 
Assuming the flare spectrum to be blackbody, this result may suggest that the temperature of it is low ($< 9000 \: \mathrm{K}$). On the other hand, the obtained flux ratio can be also explained by the Balmer and Paschen continuum flux ratio of optically thin radiation with a relatively low nonthermal electron beam of less than $5 \times 10^{11} \: \mathrm{erg \: cm^{-2} \: s^{-1}}$ (see Figure 14 and Table 6 in \cite{Brasseur_2023}).
The relationship between this value and spectral models will be discussed in detail in our future work. 

In these days, some studies have estimated the UV flux from optical flare data because it is important in terms of its effect on exoplanets (\cite{Feinstein_2020}; \cite{Howard_2020}).
On the other hand, there are studies that point to the discrepancy between such estimates and observed flux (\cite{Kowalski_2019}; \cite{Brasseur_2023}).
The fact that NUV has the clear peak before the white-light peak, as found in this study, also warns against simple estimation of UV flux from optical continuum data.
We need to more simultaneous UV and white-light samples to establish the picture of the relationship between UV and white-light flares.

\subsection{Energy Distribution} \label{sec:energy distribution}
We calculated radiated energy at each band and kinetic energy of the erupted prominence to investigate the energy distribution of this flare. We present quiescent-subtracted light curves as in Figure \ref{fig:rising_curve}a and calculated radiated energy at each band and the kinetic energy of the erupted prominence in the following manner:

 \textbf{X-ray:} We calculated X-ray fluxes at each phase from the fitting (Figure \ref{fig:all_X_spectrum_app}) using the \texttt{flux} command in \texttt{xspec}. Then, we converted fluxes to luminosity using the distance of $5.05 \: \mathrm{pc}$ \citep{Gaia_2018}. We subtracted the luminosity of Phase 0 as quiescence and created the quiescent-subtracted light curve as shown in Figure \ref{fig:rising_curve}a. We assumed that the peak and start of the X-ray and $\mathrm{H\alpha}$ flare are coincident (e.g., \cite{Kane_1974}) and fitted the decay phase with the exponential function. 
 When we fitted the X-ray light curve during the decay phase, we also assumed that the relationship between $e$-folding time of the X-ray ($\tau_{\mathrm{decay}}^{\mathrm{X-ray}}$) and $\mathrm{H\alpha}$ ($\tau_{\mathrm{decay}}^{\mathrm{H\alpha}}$):
 \begin{equation}
     \log \tau_{\mathrm{decay}}^{\mathrm{X-ray}} = \log \tau_{\mathrm{decay}}^{\mathrm{H\alpha}} + 0.1 \pm 0.6,
 \end{equation}
 which is empirically obtained in \citet{Kawai_2022}.
 Finally, we calculated radiated energy:
 \begin{eqnarray}\label{eq:flare_energy}
     E &=& \int_{\mathrm{rise}} L_{\mathrm{flare}} (t) dt + 
     \int_{\mathrm{decay}} L_{\mathrm{flare}} (t) dt \nonumber \\
       &\sim& \int_{\mathrm{rise}} (L_{\mathrm{peak}} t /\tau_{\mathrm{rise}}) dt \nonumber \\
       &+& \int_{\mathrm{decay}} L_{\mathrm{peak}} \exp\{ -(t-\tau_{\mathrm{rise}}) / \tau_{\mathrm{decay}} \} dt \nonumber \\
       &\sim& [(L_{\mathrm{peak}} \times \tau_{\mathrm{rise}})/2] + [L_{\mathrm{peak}} \times \tau_{\mathrm{decay}}],
 \end{eqnarray}
 where $L_{\mathrm{peak}}$ is the peak luminosity, $\tau_{\mathrm{rise}}$ is the time between the start and the peak of the flare, and $\tau_{\mathrm{decay}}$ is the $e$-folding time of the decay phase. 
 The 0.5$-$4 keV X-ray peak luminosity and radiated energy are derived to be $1.9 \times 10^{29} \: \mathrm{erg \: s^{-1}}$ \& $9.2 \times 10^{32} \: \mathrm{erg}$, respectively (Table \ref{tab:flare_parameters} and \ref{tab:distribution}).

  \textbf{$\mathbf{H\alpha}$:} We calculated the quiescent-subtracted equivalent widths of each time by integrating the symmetric components of $\mathrm{H\alpha}$ (c.f. Figure \ref{fig:Ha_spectrum}b). After that, we converted the equivalent widths of each time to luminosity using Equation (\ref{eq:L_EW}). We subtracted the average luminosity during two frames (at $-6 \sim 0$ min in Figure \ref{fig:rising_curve}) before the start of $\mathrm{H\alpha}$ flare as quiescence. Then, we created the quiescent-subtracted light curve (Figure \ref{fig:rising_curve} a2) and calculated the radiated energy using Equation (\ref{eq:flare_energy}).
   The $\mathrm{H\alpha}$ peak luminosity and radiated energy are derived to be $7 \times 10^{27} \: \mathrm{erg \: s^{-1}}$ and $2.7 \times 10^{31} \: \mathrm{erg}$, respectively (Table \ref{tab:flare_parameters} \& \ref{tab:distribution}).
  Note that the radiated energy of the blue-shifted excess component is not included in this energy.

 \textbf{NUV:} We calculated the flare luminosity of each time from the light curve file created by \texttt{uvotevtlc} as discussed in Section \ref{sec:rising}. We subtracted the median luminosity as quiescence of 20 minutes before the start of the NUV flare. Since we only observed the rising phase in NUV as shown in Figure \ref{fig:rising_curve}b, we calculated the lower limit of the radiation energy by integrating the observed light curve. We also assumed that $e$-folding time of NUV is shorter than that of white-light (e.g., \cite{Paudel_2021}; \cite{Brasseur_2023}; \cite{Tristan_2023}) and calculated the upper limit of the radiated energy using $e$-folding time of white-light flare and Equation (\ref{eq:flare_energy}).
   The NUV peak luminosity and radiated energy are derived to be $3.3 \times 10^{28} \: \mathrm{erg \: s^{-1}}$ and $(0.2-4.0) \times 10^{31} \: \mathrm{erg}$, respectively (Table \ref{tab:flare_parameters} \& \ref{tab:distribution}).

 \textbf{White light:} We calculated the white-light bolometric energy of the flare from the white-light light curve using the method introduced by \citet{Shibayama_2013}.
First, we divided the white-light light curve into the global trend and the flare component as shown in Figure \ref{fig:tess_energy}a. We then took the difference between them and created the detrended light curve (Figure \ref{fig:tess_energy}b). We took the flux of the detrended curve as the ratio of the luminosity of flare luminosity to that of the star ($C'_{\mathrm{flare}} (t)$) and estimated the flare area ($A_{\mathrm{flare}} (t)$) as shown in the Equation (5) of \citet{Shibayama_2013}, 
\begin{equation}
    A_{\mathrm{flare}}(t) = \frac{\pi R^{2} C'_{\mathrm{flare}}(t) \int R_{\lambda} B_{\lambda} (T_{\mathrm{eff}}) d \lambda}{\int R_{\lambda} B_{\lambda} (T_{\mathrm{flare}}) d \lambda } 
\end{equation}
where $\lambda$ is the wavelength, $B_{\lambda}$ is the Planck function, $R_{\lambda}$ is the TESS response function \citep{Ricker_2015}, $T_{\mathrm{eff}}$ is the effective temperature of the star ($3270 \: \mathrm{K}$; \cite{Paudel_2021}), $T_{\mathrm{flare}}$ is a flare temperature of $10000 \: \mathrm{K}$ (\cite{Mochnacki_1980}; \cite{Hawley_1992}), and $R$ is the radius of the star ($0.35 R_{\odot}$; \cite{Paudel_2021}). Assuming that flare radiation is a blackbody with a temperature of $T_{\mathrm{flare}} = 10000 \: \mathrm{K}$, flare luminosity $L_{\mathrm{flare}}$ is
\begin{equation}
    L_{\mathrm{flare}}(t) = \sigma_{\mathrm{SB}} T_\mathrm{flare}^4 A_{\mathrm{flare}}(t)
\end{equation}
where $\sigma_{\mathrm{SB}}$ is the Stefan-Boltzmann constant. Finally, we obtained the white-light bolometric flare energy by integrating $ L_{\mathrm{flare}}(t)$ over the duraion of the white-light flare (the green-shaded period in Figure \ref{fig:tess_energy}).
The white-light bolometric energy is derived to be $3.4 \times 10^{32} \: \mathrm{erg}$ (Table \ref{tab:distribution}).

We also calculated TESS band white-light energy by using equation (\ref{eq:L_TESS}):
\begin{eqnarray}
    E_{\mathrm{WLF}}^{\mathrm{TESS}} &=& \int L^{\mathrm{TESS}}_{\mathrm{quiescence}} \times \Delta f^{\mathrm{TESS}} (t) dt \nonumber \\
    &=& 2.4 \times 10^{32} \: \mathrm{erg}.
\end{eqnarray}

 \textbf{Kinetic Energy:} We calculated the range of the kinetic energy ($Mv_{\mathrm{blue}}^{2}/2$) of the erupted prominence using the mass range in Equation (16) and the peak velocity ($v_{\mathrm{blue}}$) shown in Figure \ref{fig:fitting_results}f. 
 The kinetic energy range is $(0.02-15) \times 10^{31} \: \mathrm{erg}$ (Table \ref{tab:distribution}).

 All parameters for the flare and the prominence are listed in Tables \ref{tab:flare_parameters} and \ref{tab:distribution}. 
 According to \citet{Ikuta_2023}, a white-light flare of this magnitude occurs once every $\sim 120$ ks on EV Lac.
 X-ray and white-light radiated energy have the same order of magnitudes and they are one order higher than NUV and $\mathrm{H\alpha}$ radiation. 
 Though there is the large uncertainties in the kinetic energy, it roughly corresponds to the value expected from the flare-kinetic energy scaling law (\cite{Takahashi_2016}; \cite{Inoue_2023}).

 Some previous studies have investigated the flare energy in X-ray and white light simultaneously (\cite{Emslie_2012}; \cite{Osten_2015}; \cite{Guarcello_2019}; \cite{Kuznetsov_2021}; \cite{Paudel_2021}; \cite{Stelzer_2022}; \cite{Namekata_2023}).
 \citet{Namekata_2023} summarized the data and showed that there is several orders of magnitude variance in the distribution of the ratio of X-ray to white-light flare energy. 
 Our obtained value ($E_{\mathrm{WLF, \: bol}} / E_{\mathrm{X}} \sim 0.4$) is in the variance. 
 As solar studies also show that there is about an order of magnitude dispersion in flare energy distribution (\cite{Emslie_2012}; \cite{Aschwanden_2017}), our result suggests the diversity of stellar flare energy distribution.

\section{Summary and Conclusion}\label{sec:Summary}
We observed EV Lac on 2022 October 24$-$27, and reported the first multiwavelength (X-ray, NUV, white light, and $\mathrm{H\alpha}$) detection of a stellar flare accompanied by $\mathrm{H\alpha}$ blue-shifts, starting at 12:28 on October 25.
The multiwavelenth observed flare is a good sample for studying the whole picture of a flare accompanied by a mass ejection. The observed flare shows the following characteristics:
\begin{enumerate}
    \item The radiation energies are $9.2 \times 10^{32} \: \mathrm{erg}$ (X-ray), $(0.2-4.0) \times 10^{31} \: \mathrm{erg}$ (NUV), $3.4 \times 10^{32} \: \mathrm{erg}$ (White-light), and $2.7 \times 10^{31} \: \mathrm{erg}$ ($\mathrm{H\alpha}$).
    \item One hour after the flare peak, a blue-shifted excess component of $\mathrm{H\alpha}$ appeared, with its Doppler velocity at $\sim 100 \: \mathrm{km \: s^{-1}}$
    \item When assuming that the observed  blue-shifted excess component is attributed to a prominence eruption, the mass and kinetic energy of the prominence are estimated to
    \begin{eqnarray} 
    1 \times 10^{15} \; \mathrm{g} < &M& < 1 \times 10^{18} \; \mathrm{g} \\
    2 \times 10^{29} \; \mathrm{erg} < &E_{\mathrm{kin}}& < 2 \times 10^{32} \; \mathrm{erg},
    \end{eqnarray}
    respectively. This follows the energy-mass scaling law of solar and stellar flares (\cite{Takahashi_2016}; \cite{Inoue_2023}; \cite{Notsu_2023}).
    \item The rising phase of the white-light flare has a substructure consisting of a gradual rise and a rapid rise. Even during the gradual rise of white-light, NUV emission has already increased rapidly. The ratio of flux in NUV to white light at the peak during the gradual phase was $\sim 0.49$.
\end{enumerate}

During this campaing observation, we have also performed coodinated observations with Five-hundred-meter Aperture Spherical radio Telescope (FAST; \cite{Nan_2006}; \cite{Nan_2011}; \cite{Zhang_2023}) and 85 cm telescope at Xinglong Station of National Astronomical Observatories, Chinese Academy of Sciences. We will further discuss the flare radiation mechanism by combining radio and multi-band optical photometric data in our upcoming papers.

\begin{ack} 
The NICER and Swift data used in this study were obtained through ToO program ID: 43723 (NICER) and ID: 17657 (Swift), respectively.
This paper includes data collected with the TESS mission, obtained from the MAST data archive at the Space Telescope Science Institute (STScI). 
Funding for the TESS mission is provided by the NASA Explorer Program.
STScI is operated by the Association of Universities for Research in Astronomy, Inc., under NASA contract NAS 5–26555.
The optical spectroscopic data used in this study were obtained through 2022B open-use program with the 2m Nayuta telescope, which is located at Nishi-Harima Astronomical Observatory, Center for Astronomy, University of Hyogo. 
We thank Megumi Shidatsu (Ehime University) for her useful comments and discussions on the UVOT data analysis.
We thank Swift helpdesk for the suggestion on the UVOT data analysis.
We thank Isaiah Tristan and Adam Kowalski (University of Colorado Boulder) for providing their modeling data.
We thank Hui Tian (Peking University) for his helpful comments and suggestions.
This work is supported by JSPS KAKENHI Grant Numbers 21J00316 (K.N.), 20K04032, 20H05643 (H.M.), 21J00106 (Y.N.), 21H01131 (H.M., D.N., S.H., K.S.), 21H04493 (T.G.T.) and RIKEN Hakubi project (PI: Teruaki Enoto).
Y.N. acknowledge support from NASA ADAP award program Number 80NSSC21K0632 (PI: Adam Kowalski). 
\end{ack}

\appendix
\begin{figure*}[] 
\begin{center}
\includegraphics[width=16cm]{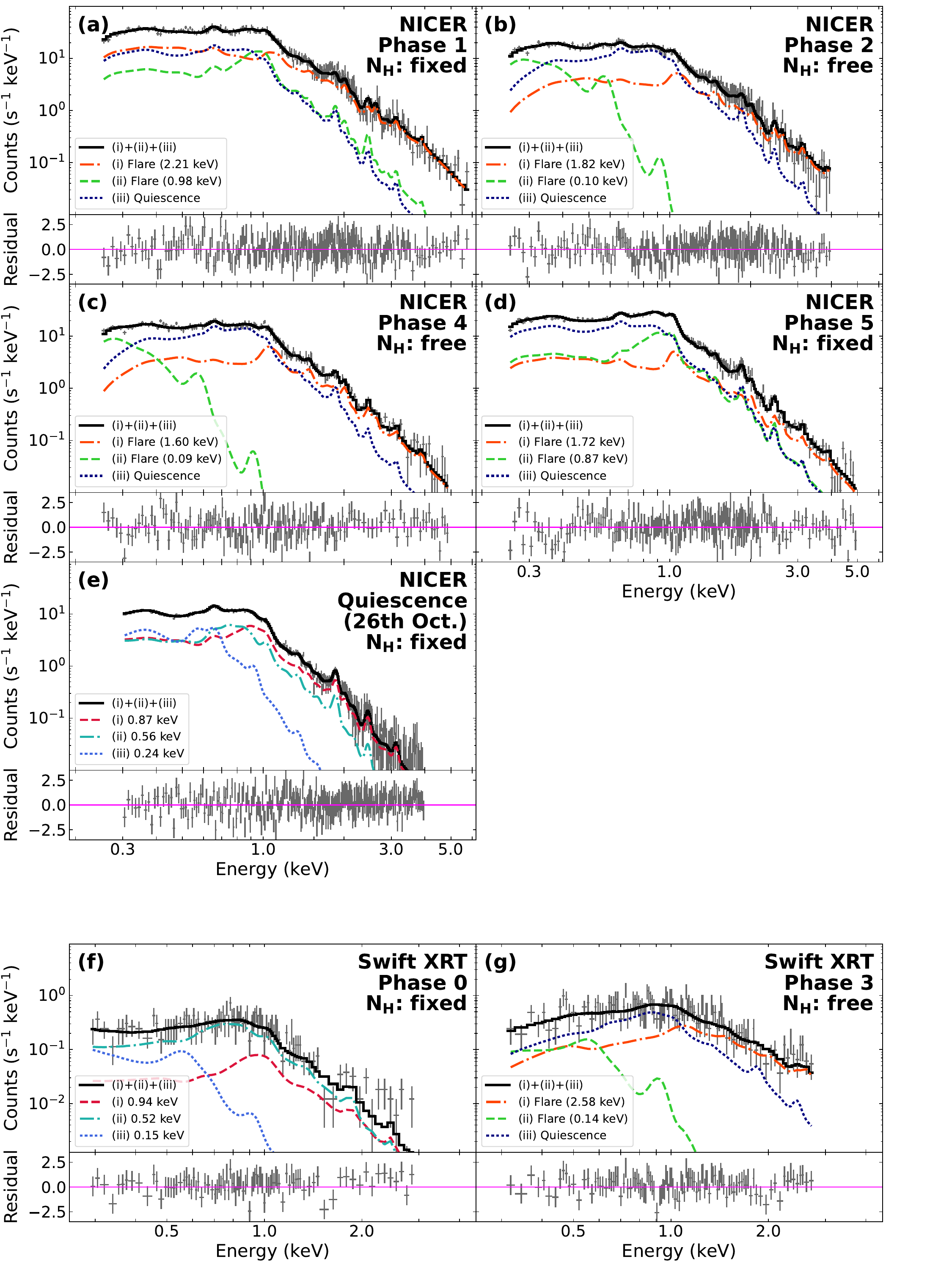} 
\end{center}
\caption{X-ray spectra at each phase shown in Figure \ref{fig:1025_curve}. Panels (a), (b), (c), (d) and (g): Spectra during the flare. The best-fit curves for the quiescence spectrum + two-component \texttt{vapec} models are shown by solid black lines for flare spectra. Navy dotted lines, orange dashdot lines, and lime dashed lines represent the quiescence spectrum, the high temperature component, and the low temperature component, respectively. Panels (e) and (f): Spectra during the quiescence. The best-fit curves for the three-component \texttt{vapec} models are shown by solid black lines for quiescence spectra. Red dashed lines, sea green dashdot lines, and cornflower blue lines represent the high temperature component, the medium temperature component, and the low temperature component,respectively.}
\label{fig:all_X_spectrum} 
\end{figure*}

\begin{table*}[t]
\caption{Best-fit parameters of spectra during the flare\footnotemark[$*$]}
\label{tab:fitting_q_2vapec}
\begin{center}
\begin{tabular}{cccccccc}
\hline
& & \multicolumn{4}{c}{NICER} & \multicolumn{1}{c}{Swift XRT} \\
& & Phase 1 & Phase 2 & Phase 4 & Phase 5 & Phase 3 \\ \hline \hline
\multicolumn{2}{c}{Exposure (ks)} & $0.42$ & $0.71$ & $0.71$ & $1.19$ & $0.82$  \\ \hline
\multicolumn{2}{c}{\texttt{tbabs}} & & & & & \\
\multicolumn{2}{c}{$N_{\mathrm{H}}$ ($10^{18}$ $\mathrm{cm^{-2}}$)} & $4.00$ & $348_{\pm 46.3}$&$354_{\pm42.2}$& $4.00$ & $925_{\pm 334}$ \\ \hline
\multicolumn{2}{c}{\texttt{vapec} (High Temp.)} & & & & & \\
\multirow{2}{*}{Temperature} &$kT$ (keV)& $2.21_{\pm0.13}$ & $1.82_{\pm0.14}$ & $1.60_{\pm0.07}$ & $1.72_{\pm0.11}$ & $2.58_{\pm0.91}$ \\
&$T$ (MK)& $25.7_{\pm1.51}$ & $21.1_{\pm1.63}$ & $18.6_{\pm0.81}$ & $20.0_{\pm1.28}$ & $29.9_{\pm10.6}$ \\
\multicolumn{2}{c}{norm ($10^{-2}$)} & $2.09_{\pm0.12}$ & $0.68_{\pm0.46}$ & $0.64_{\pm0.04}$ & $0.42_{\pm0.04}$ & $1.03_{\pm0.12}$ \\ \hline
\multicolumn{2}{c}{\texttt{vapec} (Low Temp.)} & & & & & \\
\multirow{2}{*}{Temperature}&$kT$ (keV)& $0.98_{\pm0.04}$ & $0.10_{\pm0.01}$ & $0.09_{\pm0.01}$ & $0.87_{\pm0.03}$ & $0.14_{\pm0.06}$ \\
&$T$ (MK)& $11.37_{\pm0.46}$ & $1.16_{\pm0.12}$ & $1.05_{\pm0.12}$ & $10.11_{\pm0.35}$ & $1.63_{\pm0.70}$ \\
\multicolumn{2}{c}{norm ($10^{-2}$)} & $0.61_{\pm0.11}$ & $1.35_{\pm0.42}$ & $1.82_{\pm0.83}$ & $0.40_{\pm0.04}$ & $1.06_{\pm1.36}$ \\ \hline
\multirow{13}{*}{$Z$ ($Z_{\odot}$)} & He & $1.00$ & $1.00$& $1.00$ & $1.00$& $1.00$ \\
& C & $0.74_{\pm0.18}$ & $0.29_{\pm0.09}$& $0.17_{\pm0.06}$& $0.56_{\pm0.14}$& 0.29\\
& N & $0.74_{\pm0.18}$ &$0.29_{\pm0.09}$ &$0.17_{\pm0.06}$ & $0.56_{\pm0.14}$& 0.29 \\
& O & $0.74_{\pm0.18}$ &$0.29_{\pm0.09}$ & $0.17_{\pm0.06}$& $0.56_{\pm0.14}$& 0.29 \\
& Ne & $1.00$ & $1.00$& $1.00$& $3.16_{\pm0.42}$& $1.00$ \\
& Mg & $0.43_{\pm0.20}$ & $1.00$& $1.00$&$1.00$ & $1.00$ \\
& Al & $1.00$ & $1.00$ & $1.00$& $1.00$& $1.00$ \\
& Si & $0.65_{\pm0.14}$ & $0.49_{\pm0.23}$& $0.79_{\pm0.21}$& $0.68_{\pm0.11}$& 0.49 \\
& S & $1.00$ & $1.00$& $1.00$& $1.00$ & $1.00$ \\
& Ar & $1.00$ & $1.00$& $1.00$& $1.00$& $1.00$ \\
& Ca & $1.00$ & $1.00$& $1.00$& $1.00$& $1.00$ \\
& Fe & $0.30_{\pm0.04}$ & $0.30_{\pm0.10}$& $0.32_{\pm0.08}$& $0.29_{\pm0.03}$& 0.30 \\
& Ni & $1.00$ & $1.00$& $1.00$& $1.00$& $1.00$ \\ \hline
\multicolumn{2}{c}{$\chi^{2}$ ($d.o.f$)} & 268 (250) & 215 (219) & 211 (160)& 252 (203) & 86 (106) \\ \hline
\end{tabular}
\end{center}
\begin{tabnote}
    \footnotemark[$*$] The error ranges correspond to $90\%$ confidence level. Values without errors mean that they are fixed. \\
    % For Phase 3 spectrum, elemental abundances were fixed to the best-fit value of Phase 2.
\end{tabnote}
\end{table*}

\section*{X-ray spectral fitting with the fixed quiescent component} \label{sec:without_quiescence}
We also conducted X-ray spectral fitting with the fixed quiescent component during the flare.
This fitting method assumes that there is no significant variation in the quiescent component during the flare (e.g., \cite{Hamaguchi_2023}).

Based on the results of the spectral analysis of quiescent spectra (c.g. Figure \ref{fig:all_X_spectrum_app} e and f), we analyzed the spectra for each phase after the flare occurred (Phase 1-5). For the analysis of flare spectra, we fixed the best fit values of October 26 NICER spectrum as the quiescent component. Then, we fit the flare component with two temperature collisionally-ionized models (\texttt{vapec}).
We linked abundance between flare components.
For Phase 3 spectrum obtained by Swift XRT, we fixed abundance to the best-fit value of Phase 2, which is the closet phase to Phase 3, since it cannot be well determined due to a lack of statistics.
As in the analysis of quiescent spectra, we tried to fit flare spectra with the model that multiplies the sum of the quiescent and flare components by an interstellar absorption fixed to the literature value ($N_{\mathrm{H}} = 4.0 \times 10^{18} \: \mathrm{cm}^{-2}$; \cite{Paudel_2021}). However, spectra of Phase 2, 3, and 4 were not acceptably explained by this model. Specifically, the temperature of the flare component became overwhelmingly higher ($3-70 \: \mathrm{keV}$ / $35-810 \: \mathrm{MK}$) than Phase 1, which is closer to the peak of the flare, to fit the low energy side ($< 1 \: \mathrm{keV}$) of Phase 2-4 spectra. These fitting parameters are physically unacceptable because the loop plasma should be cooled via radiation after the peak of the flare \citep{Shibata_2011}. 
As mentioned in Section \ref{sec:light_curve}, some small flares occurred before and after Phase 3 and 4. Therefore, we tried to fit the flare component of Phase 3 and 4 by adding an extra \texttt{vapec} components with a fixed interstellar absorption. However, the spectra could not be fitted well no matter how many components were added.

Then, to improve the fits, we set $N_{\mathrm{H}}$ free for Phase 2, 3, and 4.
We unified the number of flare \texttt{vapec} components to two for all flare spectra. When we set $N_{\mathrm{H}}$ free, spectra of Phase 2, 3, and 4 could be well fitted.
These flare spectra and the results of the spectral fit are shown in Figure \ref{fig:all_X_spectrum} a-d,  and g. Table \ref{tab:fitting_q_2vapec} lists the best-fit parameters for these fitting. As shown in Table \ref{tab:fitting_q_2vapec}, best-fit values of hydrogen column density increased by two orders of magnitude from $4\times10^{18} \: \mathrm{cm}^{-2}$ to $3.5-9.3\times10^{21} \: \mathrm{cm}^{-2}$ during Phase 2-4.

\citet{Moschou_2017} suggested that CMEs passing through the line-of sight direction cause increase of hydrogen column density. In this flare, the ejected plasma, seen as the blueshift of $\mathrm{H\alpha}$, may also have caused X-ray absorption. 
When the prominence is optically thick, from Equations (\ref{eq:ne_value}), (\ref{eq:ne_nh_ratio}), and (\ref{eq:ne2D31}), hydrogen column density of the prominence $N_{\mathrm{H}}^{\mathrm{p}} = n_{\mathrm{H}} D$ is
\begin{equation}
    7 \times 10^{19} \: \mathrm{cm^{-2}} < N_{\mathrm{H}}^{\mathrm{p}} < 6 \times 10^{21} \: \mathrm{cm^{-2}}.
\end{equation}
This value is consistent with the observed hydrogen column density during Phase 2-4 shown in Table \ref{tab:fitting_q_2vapec}.

However, the fitting parameters of low temperature components during Phase 2-4 (0.10 keV / 0.14 keV / 0.09 keV) are lower than the lowest component of the quiescence (0.24 keV). This means that a part of the flare loop is much cooler than the coronal plasma in the quiescent phase. As such situation is not natural, we hesitate to declare the increase of hydrogen column density and adopted the fitting without the fixed quiescent component as described in Section \ref{sec:fitting_X-ray}.

\end{document}